\documentclass[journal]{IEEEtran}
\usepackage{graphicx}
\usepackage{subcaption}
\usepackage{float}
\usepackage{physics}
\usepackage{cite}
\usepackage{amsmath,amsthm,amsfonts,amssymb,amscd}
\usepackage{amsmath}
\usepackage{xcolor}
\usepackage{bm}
\usepackage{cuted}
\usepackage{algorithm}
\usepackage[noend]{algpseudocode}
\usepackage{multirow}

\newcommand{\vectext}{\text{vec}}

\begin{document}
\title{Transmission Line Parameter Estimation Under Non-Gaussian Measurement Noise}
%
%
%

\author{Antos~Cheeramban~Varghese,~\IEEEmembership{ Student Member,~IEEE,}
        Anamitra~Pal,~\IEEEmembership{Senior~Member,~IEEE,}
        and~Gautam~Dasarathy,~\IEEEmembership{Senior Member,~IEEE}
\thanks{This work was supported in part by the National Science Foundation
(NSF) Grants OAC 1934766, CCF 2048223, and ECCS 2145063.\\
The authors are associated with the School of Electrical, Computer, and
Energy Engineering, Arizona State University, Tempe, AZ 85287,
USA.}
}

%
%

\markboth{}%
{Shell \MakeLowercase{\textit{et al.}}: Bare Demo of IEEEtran.cls for IEEE Journals}
%



\maketitle

\begin{abstract}
 Accurate knowledge of transmission line parameters is essential for a variety of power system monitoring, protection, and control applications. The use of phasor measurement unit (PMU) data for transmission line parameter estimation (TLPE) is well-documented. 
 However, existing literature on PMU-based TLPE implicitly assumes the measurement noise to be Gaussian.
Recently, it has been shown that the noise in PMU measurements (especially in the current phasors) is better represented by Gaussian mixture models (GMMs), i.e., the noises are non-Gaussian.
 We present a novel 
 approach for TLPE
that can handle non-Gaussian noise in the PMU measurements. 
 The measurement noise is expressed as a GMM, whose components are identified using the expectation-maximization (EM) algorithm. Subsequently, noise and parameter estimation is carried out by solving a maximum likelihood estimation problem iteratively until convergence.
The superior performance of the proposed approach
 over 
traditional approaches such as least squares and total least squares as well as the more recently proposed minimum total error entropy approach is demonstrated by performing simulations using the 
IEEE 118-bus system as well as proprietary PMU data 
obtained from a U.S. power utility.
 
\end{abstract}


%
\IEEEpeerreviewmaketitle



\section{Introduction}
\label{Intro}
%
%
%
%
\IEEEPARstart{A}{ccurate} knowledge of transmission line parameters is critical to the success of 
power system 
applications such as
state estimation, optimal power flow, dynamic line rating, protection relay settings, and post-event fault location 
\cite{ritzmann2017novel,dasgupta_line_2013_1,shi2011transmission1}. 
However, the transmission line parameters change because of variations in temperature, humidity, and operating conditions (in the short-term), as well as due to aging and structural changes associated with sag, addition of new equipment, and reconductoring (in the long-term) \cite{shi2011transmission1, ritzmann_method_2016, du2012line1, lin2019synchrophasor}.
As such, 
the line parameters
must be estimated periodically (e.g., every few months \cite{asprou2017identification}).
The use of phasor measurement unit (PMU) data for transmission line parameter estimation (TLPE) has particularly found prominence as the numbers of PMUs have grown in the transmission system \cite{ritzmann2017novel,shi2011transmission1,dasgupta_line_2013_1, xue2019linear, wehenkel2020parameter, gupta2021compound}.

PMU-based TLPE is a linear regression problem in which the regressand (henceforth called the \textit{dependent variable})
and the regressor (henceforth called the \textit{independent variable}) 
are composed of PMU measurements. When Gaussian noise is present in the dependent variables and the independent variables are noise-free, the optimal solution to the linear regression problem is obtained using the least squares (LS) method. 
When Gaussian noise is present in both the dependent and the independent variables, the total least squares (TLS) method is employed to obtain the optimal solution.
However, prior research has shown that the noise in PMU measurements is non-Gaussian \cite{wang2017assessing,ahmad2019statistical} (see Appendix \ref{AppendixA} for a detailed explanation regarding the nature and source of noise in PMU measurements).
It has also been demonstrated that the 
performance of LS and TLS
degrade when non-Gaussian noise is present in the dependent and/or independent variables \cite{wang2017maximum}. The research focus of this paper is to perform linear regression in presence of non-Gaussian noise in both the
variables.

The primary challenge in handling non-Gaussian noise in linear regression problems is that as the noise may not have an analytically tractable probability density function (PDF), a closed-form solution to the problem may be difficult (or even impossible) to obtain. 
One way in which this problem can be tackled is by approximating the non-Gaussian distribution 
using a finite weighted sum of known Gaussian densities, called a Gaussian mixture model (GMM) \cite{zhao2017framework, goodfellow2016deep} (see Appendix \ref{AppendixB} for a mathematical explanation of how an arbitrary distribution can be approximated by a GMM).
In practice, however, one does not know the Gaussian densities \textit{a priori}. Therefore, in this paper we formulate a joint parameter and noise estimation problem, where in addition to estimating the parameters, the characteristics of the non-Gaussian noise, expressed as GMMs, are also estimated. 
The knowledge about the noise characteristics of a PMU's measurements provides the additional advantage that by tracking their variations, one can decide the right time to calibrate the 
PMU \cite{pal2015online}.

The key contributions of this paper are as follows:
\begin{itemize}
    \item A novel technique to optimally estimate parameters of linear regression problems in which the dependent variables have non-Gaussian noise.
    \item The technique is robust enough to estimate the characteristics of the GMM that approximates the non-Gaussian noise.
    \item The technique is extended to solve an errors-in-variables (EIV) problem in which both the dependent and the independent variables have non-Gaussian noise.
    \item Successfully estimating line parameters of the IEEE 118-bus system and an actual power system in presence of Gaussian/non-Gaussian noise in the PMU measurements.
\end{itemize}

\section{State-of-the-art}
\label{SOTA}

Prior research on TLPE can be grouped into two categories. The first category considered noise in only the dependent variables (e.g., \cite{gurusinghe2017efficient,xue2019linear,milojevic2018utilization_1}), while the second category considered noise in both the dependent and the independent variables (e.g., \cite{ding2011transmission_1,goklani2020instrument_1,ritzmann_method_2016,mishra2015kalman_1,yu2017patopa_1,lin2019synchrophasor,singh2018medium,wehenkel2020parameter,gupta2021compound}).
Ref. \cite{xue2019linear} studied the effect of phase angle difference errors on line parameter calculations. An LS-based line parameter estimation technique was proposed in \cite{gurusinghe2017efficient} to handle noise in dependent variables. A robust-M estimator for three-phase line parameter estimation was proposed in \cite{milojevic2018utilization_1}.
Since TLPE requires the independent variables to also be composed of PMU measurements, the assumption of these variables being noise-free, limits the accuracy of the algorithms falling under the first category.
 
The algorithms proposed in the papers belonging to the second category considered noise in both dependent and independent variables.
Ritzmann et al. proposed a set of correction constants to represent the noise in voltage and current measurements obtained from PMUs \cite{ritzmann_method_2016}. 
A recursive LS-based method for line parameter estimation with focus on energy management was proposed in \cite{lin2019synchrophasor}. 
Weighted TLS, ordinary LS, and ordinary TLS were used for estimating line parameters of a three-phase transmission line in \cite{wehenkel2020parameter}.
A combination of clustering and TLS was used for three-phase admittance matrix estimation in \cite{gupta2021compound}. 
Ding et al. proposed a TLS-based parameter estimation method in which both dependent and independent variables were noisy \cite{ding2011transmission_1}. 
A method for calibrating instrument transformers and simultaneously estimating three-phase line parameters considering noise in both sets of variables was proposed in   \cite{goklani2020instrument_1}. 
A recursive regression based on Kalman filtering was used for estimating the three-phase line parameters in \cite{mishra2015kalman_1}. 
Ref. \cite{yu2017patopa_1} presented a joint parameter and topology estimation framework using maximum likelihood estimation. 
Line parameter estimation using an LS-based method and focusing on dynamic line rating was discussed in \cite{singh2018medium}.  
However, all the above-mentioned papers assumed Gaussian noise in the PMU measurements, which has been disproved recently \cite{wang2017assessing,ahmad2019statistical}.
To the best of the authors' knowledge, this is the first paper that accounts for the presence of non-Gaussian noise in PMU measurements when performing PMU-based TLPE.

\section{Problem Formulation}
\label{ProbForm}

\subsection{TLPE Model and its Abstraction}
\label{TLPEAbstraction}

\begin{figure}[H]
\centering
\includegraphics[width=0.38\textwidth]{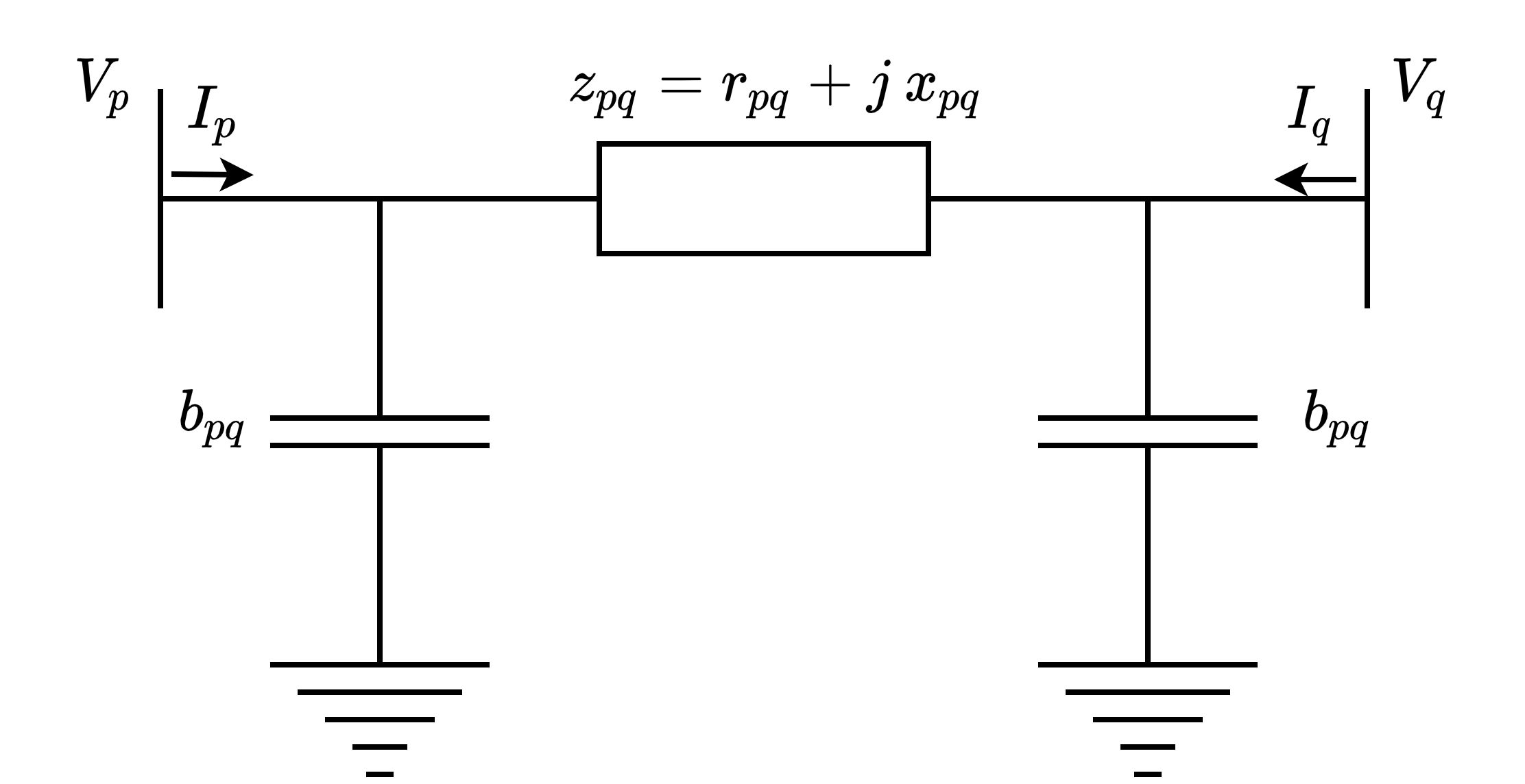}
\caption{$\pi$-model of a medium length transmission line}
\label{fig:medium_line}
\end{figure}

A medium-length positive sequence transmission line model has been considered for the analysis conducted here. However, with appropriate modifications, the proposed formulation can also be applied to other line models.
The line segment is modeled as a $\pi$-section (see Fig. \ref{fig:medium_line}), where $p$ and $q$ denote its sending and receiving ends, respectively.

Applying Kirchhoff's circuit laws to the transmission line model shown in Fig. \ref{fig:medium_line}, the following equation can be written.  
\begin{equation}
\label{eqn:TLPE_Basic}
         \begin{aligned}
         I_p &= b_{pq} V_p + (V_p - V_q) y_{pq}\\
         I_q &= b_{pq} V_q - (V_p - V_q) y_{pq}.
         \end{aligned}
\end{equation}
 In \eqref{eqn:TLPE_Basic}, $I$ and $V$ denote the complex current and voltage measurements obtained from PMUs placed at the two ends of the line (\textcolor{black}{represented by $p$ and $q$}), 
  \textcolor{black}{$b_{pq} \in \mathbb{R}$} denotes the shunt admittance (susceptance) present at each end, and  \textcolor{black}{$y_{pq} \in \mathbb{C}$} denotes the series admittance, which is inverse of the series impedance,  \textcolor{black}{$z_{pq} \in \mathbb{C}$}. 
 The real and imaginary parts of $z_{pq}$ are the resistance, \textcolor{black}{$r_{pq} \in \mathbb{R}$}, and reactance, \textcolor{black}{$x_{pq} \in \mathbb{R}$}, of the line. 
The objective of this work is to estimate $r_{pq}$, $x_{pq}$, and $b_{pq}$ from the PMU measurements.
 Expressing the complex currents, voltages, and line parameters of \eqref{eqn:TLPE_Basic} in their Cartesian form and rearranging the terms, we get



   \begin{gather}
   \label{eqn:TLPE1a}
 \begin{bmatrix}  I_{p_r} \\  I_{p_i} \\  I_{q_r} \\  I_{q_i}\end{bmatrix}
 =  \begin{bmatrix}  -V_{p_i} &  ( V_{p_r}- V_{q_r}) & - (  V_{p_i} -  V_{q_i}) \\  V_{p_r} &  ( V_{p_i}- V_{q_i}) & ( V_{p_r}- V_{q_r}) \\
 -V_{q_i} &  -( V_{p_r}- V_{q_r}) &  (  V_{p_i} -  V_{q_i}) \\  V_{q_r} &  -( V_{p_i}- V_{q_i}) & -( V_{p_r}- V_{q_r})\end{bmatrix} 
 \begin{bmatrix}b_{pq} \\ y_{pq_r} \\ y_{pq_i} \end{bmatrix}
 \end{gather}
\textcolor{black}{where, $y_{pq_r} \in \mathbb{R}$ and $y_{pq_i} \in \mathbb{R}$ denote the real and imaginary parts of $y_{pq}$.}
It can be observed from \eqref{eqn:TLPE1a} that there is a summation of voltages in the right-hand side. 
This
is a concern when non-Gaussian noise is present in the voltage measurements because it might be more difficult to approximate summation of two non-Gaussian noises by a GMM in comparison to approximating the individual noises (by GMMs). 
To alleviate this concern, linear algebra is used to express the voltages independently. 
\textcolor{black}{This results in the following equation, where $t$ denotes the time instant at which the phasor measurements are taken.}



  \begin{gather}
   \label{eqn:TLPE_I_VY_t}
 \begin{bmatrix}   I_{p_r} (t) \\ I_{p_i}  (t) \\  I_{q_r}  (t) \\ I_{q_i}  (t) \end{bmatrix}
 =  \begin{bmatrix}   V_{p_r}  (t) &  V_{p_i}  (t) & V_{q_r}  (t)  &  V_{q_i}  (t) \\  V_{p_i}  (t) &  -V_{p_r}  (t) & V_{q_i} (t) &  -V_{q_r}  (t) \\  V_{q_r}  (t) &  V_{q_i}  (t) & V_{p_r}  (t)  &  V_{p_i}  (t) \\ V_{q_i}  (t) &  -V_{q_r}  (t) & V_{p_i} (t) &  -V_{p_r}  (t)\end{bmatrix} 
\begin{bmatrix}  Y_1 \\ Y_2 \\ Y_3 \\Y_4 \end{bmatrix}
 \end{gather}
 
In \eqref{eqn:TLPE_I_VY_t}, $Y_1 = y_{pq_r}$ \textcolor{black}{$ \in \mathbb{R}$}, $Y_2 = -(b_{pq}+ y_{pq_i})$ \textcolor{black}{$ \in \mathbb{R}$}, $Y_3 = -y_{pq_r}$ \textcolor{black}{$ \in \mathbb{R}$}, and $Y_4 = y_{pq_i}$ \textcolor{black}{$ \in \mathbb{R}$}, respectively.
\textcolor{black}{Once $\begin{bmatrix}  Y_1 & Y_2 & Y_3 & Y_4 \end{bmatrix}^T$ are estimated,
the line parameters, namely, $r_{pq}$, $x_{pq}$, and $b_{pq}$, can be calculated using \eqref{eq4}.}
   \begin{equation}
 \begin{aligned}
 \label{eq4}
   r_{pq} &= \frac{2(Y_1 - Y_3)}{(Y_1 - Y_3)^2 + (2Y_4)^2}\\
    x_{pq} &= \frac{-4Y_4}{(Y_1 - Y_3)^2 + (2Y_4)^2}\\
    b_{pq} &= -(Y_2+Y_4).
 \end{aligned}
 \end{equation}
 
\textcolor{black}{However, in presence of noisy data, measurements obtained at a single time instant are not able to give accurate estimates of $\begin{bmatrix}  Y_1 & Y_2 & Y_3 & Y_4 \end{bmatrix}^T$.
Therefore, the
goal is to estimate the optimal values of the line parameters from noisy measurements available from distinct time instants.} 
Based on 
\eqref{eqn:TLPE_I_VY_t}, the measurements from $s$ time instants can be concatenated to obtain a system of equation as
   \begin{gather}
   \label{eqn:TLPE_I_VY_N_time_steps}
 \begin{bmatrix}   I_{p_r} (1) \\ I_{p_i}  (1) \\  I_{q_r}  (1) \\ I_{q_i} (1)\\ \vdots \\    I_{p_r} (s) \\ I_{p_i}  (s) \\  I_{q_r}  (s) \\ I_{q_i}  (s)  \end{bmatrix}
 =  \begin{bmatrix}   V_{p_r}  (1) &  V_{p_i}  (1) & V_{q_r}  (1)  &  V_{q_i}  (1) \\  V_{p_i}  (1) &  -V_{p_r}  (1) & V_{q_i} (1) &  -V_{q_r}  (1) \\  V_{q_r}  (1) &  V_{q_i}  (1) & V_{p_r}  (1)  &  V_{p_i}  (1) \\ V_{q_i}  (1) &  -V_{q_r}  (1) & V_{p_i} (1) &  -V_{p_r}  (1) \\
 \vdots &  \vdots &  \vdots &  \vdots \\  V_{p_r}  (s) &  V_{p_i}  (s) & V_{q_r}  (s)  &  V_{q_i}  (s) \\  V_{p_i}  (s) &  -V_{p_r}  (s) & V_{q_i} (s) &  -V_{q_r}  (s) \\  V_{q_r}  (s) &  V_{q_i}  (s) & V_{p_r}  (s)  &  V_{p_i}  (s) \\ V_{q_i}  (s) &  -V_{q_r}  (s) & V_{p_i} (s) &  -V_{p_r}  (s) \end{bmatrix} 
\begin{bmatrix}  Y_1 \\ Y_2 \\ Y_3 \\Y_4 \end{bmatrix}.
 \end{gather}
 
\textcolor{black}{Essentially, \eqref{eqn:TLPE_I_VY_N_time_steps} takes advantage of the fact that the line parameters change at a much slower rate than the speed at which the PMU measurements become available \cite{chatterjee2018error}.}
Equation \eqref{eqn:TLPE_I_VY_N_time_steps} can now be expressed as a classical parameter estimation problem of the form shown below
\begin{equation}
\label{linear_system_relationship}
    c^* = D^* x^*
\end{equation}
where, $c \in \mathbb{R}^{n \times 1}$ denotes the dependent variables (current measurements), $D \in \mathbb{R}^{n \times p}$ denotes the independent variables (voltage measurements), and $x \in \mathbb{R}^{p \times 1}$ is the unknown parameter 
to be estimated (function of line parameters); the symbol $^*$ in the superscript indicates \textit{true} values. 
It can be realized from \eqref{eqn:TLPE_I_VY_N_time_steps} and \eqref{linear_system_relationship}, that $n=4s$ and $p$, the number of parameters to be estimated, is $4$.

\subsection{Noise Modeling}
\label{NoiseModel}

In presence of Gaussian noise in PMU measurements, two scenarios arise. If the noise in the voltage measurements (independent variables) is ignored (i.e., Gaussian noise is only present in the current measurements (dependent variables)), the optimal estimate of the parameters can be obtained using LS. 
This scenario can be mathematically represented as 

 \begin{equation}
 \label{LSeq}
 \begin{aligned}
      c &= c^* + c_{e}\\
      D &= D^*,
 \end{aligned}
 \end{equation}

where, \textcolor{black}{$c_e \in \mathbb{R}^{n \times 1}$} is the noise in $c$. For the linear system described by \eqref{linear_system_relationship} and \eqref{LSeq}, the LS parameter estimate, $\hat{x}_{LS}$, is given by
\begin{equation}
\label{least_square_estimate}
   \hat{x}_{LS} = \left( D^T D\right)^{-1}  \left(D^T c\right)
\end{equation}

In the second scenario, Gaussian noise is present in both the dependent as well as the independent variables, and the optimal parameter estimate is obtained using TLS.
This scenario can be mathematically represented as

 \begin{equation}
 \label{EIV measurement model}
 \begin{aligned}
       c &= c^* + c_e\\
       D &= D^* + D_e,
 \end{aligned}
 \end{equation}
 
where, \textcolor{black}{$D_e \in \mathbb{R}^{n \times p}$} is the noise in $D$. 
To obtain the TLS solution of the linear system described by \eqref{linear_system_relationship} and \eqref{EIV measurement model}, the concatenated noisy measurement matrix, 
$[ D~~c ]$, is factorized using singular value decomposition (SVD) into a matrix of singular values and left and right singular vectors.
Then, using the Eckhart-Young Theorem, the TLS solution can be obtained as \cite{markovsky2007overview}
\begin{equation}
    \hat{x}_{TLS} = v_{qq}^{-1} v_{pq}
\end{equation}
where, $v_{pq}$ is the vector of first $p$ elements and $v_{qq}$ is the $(p+1)^{th}$ element, respectively, of the $(p+1)^{th}$ column of the matrix of right singular vectors of the SVD of $[ D~~c ]$.

In this paper, we investigate both of the scenarios mentioned above under the condition that the PMU measurements have non-Gaussian noise. That is, in the analysis conducted henceforth, $c_e$ and $D_e$ will have non-Gaussian distributions.
The proposed approach for solving the resulting problems is composed of two steps: a \textit{noise estimation step} and a \textit{parameter estimation step}. In the noise estimation step, the non-Gaussian noise is approximated by an appropriate GMM. The GMM parameters are obtained using \textit{expectation-maximization} (EM). More details about this step are provided in Section \ref{NoiEst}. The parameter estimation step is explained for the first scenario in Section \ref{ParEstDep} and for the second scenario in Section \ref{ParEstInd}.

\subsection{Noise Estimation}
\label{NoiEst}
We use $c_e$ to 
demonstrate the noise estimation step.
If an $m$ component GMM is suitable for approximating $c_e$, then the following equation holds true.
\begin{equation}
    \mathbb{P}(c_e; \theta) = \sum\limits_{g=1}^m w_g \mathcal{N}(c_{e}; \mu_g, \Sigma_g).
    \label{GMM-PDF}
\end{equation}

In \eqref{GMM-PDF}, 
$w_g \geq 0$, $\sum\limits_{g=1}^m w_g = 1$, and $\mathcal{N}(c_e ; \mu, \Sigma) = \frac{1}{\sqrt{(2 \pi)^d |\Sigma|}} \\ exp(-\frac{1}{2}(c_e - \mu)^T \Sigma^{-1} (c_e-\mu))$. 
Here, $w_g, \mu_g$, and $\Sigma_g$ denotes the weight, mean vector, and covariance matrix of the $g^{th}$ Gaussian component of the GMM, respectively. These three variables are known as the GMM parameters. 
A set of these GMM parameters for the $g^{th}$ Gaussian component is denoted by $\phi_g \triangleq \{ w_g, \mu_g, \Sigma_g\}$, and the entire set of GMM parameters to be estimated is denoted by $\theta = \{\phi_g\}_{g=1}^m$. Ideally, the optimal GMM parameters can be estimated by maximizing the log-likelihood of  \eqref{GMM-PDF} with respect to the GMM parameters.
However, 
the log-likelihood has summation inside the logarithm, which makes the problem of directly maximizing the log-likelihood hard. EM overcomes this problem by introducing a new variable, $z$, and maximizing the complete data log-likelihood, as explained below.

Let the $n$ independent and identically distributed samples of $c_e$ 
be denoted by $\mathcal{S}$, i.e., $\mathcal{S} = \{c_{e_1}, c_{e_2}, \dots, c_{e_n} \}$. The identity of the Gaussian component to which each of these data points 
belong 
(out of $m$ possible Gaussian components) 
is defined as the cluster membership of those data points. Let $z=[z_1, z_2, \dots, z_m]^T$ be an $m$-dimensional binary random variable.
The $m$ possible values that $z$ can take can be denoted by $\{\zeta_g\}_{g=1}^m$, where $\zeta_g$ denotes an $m$-dimensional vector of zeros with one in the $g^{th}$ place.
The probability of $z$ taking value $\zeta_g$ is $w_g$. In other words, $\mathbb{P}(z = \zeta_g) = w_g$. Thus, the distribution of $z$ can be written as
\begin{equation}
    \mathbb{P}(z) = \prod\limits_{g=1}^m w_g^{z_g}.
\end{equation}

The joint distribution of $c_e$ and $z$ can now be expressed as
\begin{equation}
    \begin{aligned}
       \mathbb{P}(c_{e}, z) = \mathbb{P}(c_{e}) \mathbb{P}(c_{e}|z)  = \prod\limits_{g=1}^m w_g^{z_g} \mathcal{N}(c_e; \mu_g, \Sigma_g)^{z_g},
    \end{aligned}
\end{equation}
while the conditional probability of $z$ given $c_e$ is given by
\begin{equation}
    \begin{aligned}
      \mathbb{P}(z_g =1|c_{e}) &= \mathbb{P}(z = \zeta_g|c_{e})\\
      &= \frac{ \mathbb{P}(z = \zeta_g)\mathbb{P}(c_{e}|z = \zeta_g)}{\sum\limits_{g=1}^m  \mathbb{P}(z = \zeta_g) \mathbb{P}(c_{e}|z = \zeta_g)}\\
      &= \frac{w_g \mathcal{N}(c_e; \mu_g, \Sigma_g)}{\sum\limits_{g^{\prime}=1}^m w_{g^{\prime}} \mathcal{N}(c_e; \mu_{g^{\prime}}, \Sigma_{g^{\prime}})}.
    \end{aligned}
\end{equation}

Let $\mathcal{S}_c = \{(c_{e_i}, z_i)\}_{i=1}^n$. 
Now, EM maximizes 
$\mathcal{S}_c$
instead of 
$\mathcal{S}$, where the likelihood of $\mathcal{S}_c$
can be written as \cite{sahu2020new}

\begin{equation}
\label{GMM-complete-data-log-likelihood}
    \begin{aligned}
       L_c(\theta;\mathcal{S}_c ) = \prod\limits_{i=1}^n \mathbb{P}(c_{e_i}, z_i) &= \prod\limits_{i=1}^n \prod\limits_{g=1}^m w_g^{z_g^i} \mathcal{N}(c_e; \mu_g, \Sigma_g)^{z_g^i}\\
       &= \prod\limits_{i=1}^n \prod\limits_{g=1}^m (w_g \mathcal{N}(c_e; \mu_g, \Sigma_g))^{z_g^i}.
    \end{aligned}
\end{equation}

In \eqref{GMM-complete-data-log-likelihood}, $z_g^i$ represents $g^{th}$element of $z$ for the $i^{th}$ data sample. 
The complete data log-likelihood can be obtained by taking logarithm on both sides of \eqref{GMM-complete-data-log-likelihood}, as shown below.
\begin{equation}
\label{Complete log-likelihood}
    \begin{aligned}
       l_c(\theta;\mathcal{S}_c ) = \sum\limits_{i=1}^n \sum\limits_{g=1}^m z_g^i \log(w_g \mathcal{N}(c_e; \mu_g, \Sigma_g)). 
    \end{aligned}
\end{equation}

From \eqref{Complete log-likelihood},
EM can learn the GMM parameters for a given noise vector. EM has two main steps - expectation and maximization - which are repeated until convergence. The expectation step computes the conditional expectation of the complete data log-likelihood using given GMM parameters. In the maximization step, this conditional expectation of the complete data log-likelihood is maximized to obtain updated GMM parameters. 
After EM converges, the noise vector is clustered into different bins based on the cluster membership.
These three processes are described below.

\subsubsection{Expectation step}
The conditional expectation of the complete data log-likelihood can be defined as
\begin{equation}
\label{Q_definition}
    Q(\theta| \theta^{(j)}) = \mathbb{E}[l_c(\theta; \mathcal{S}_c)|\mathcal{S}, \theta^{(j)}]
\end{equation}
where $t$ denotes the iteration number, $\theta^{(j)}$ denotes the GMM parameter values at iteration $j$, and $Q$ is the auxiliary function \cite{murphy2012machine}. Substituting \eqref{Complete log-likelihood} in \eqref{Q_definition}, the conditional expectation of the complete data log-likelihood can be expressed as

\begin{equation}
\label{CECDLL-Derivation}
    \begin{aligned}
       Q(\theta &| \theta^{(j)}) = \mathbb{E}[l_c(\theta; S_c)| S,\theta^{(j)}] \\
        &= \sum\limits_{i=1}^n \sum\limits_{g=1}^m \mathbb{E}[z_g^i| S,\theta^{(j)}] \log(w_g^{(j)}   \mathcal{N}(c_{e_i}; \mu_g^{(j)}, \Sigma_g^{(j)})) \\
         &= \sum\limits_{i=1}^n \sum\limits_{g=1}^m \mathbb{P}[z_g^i = 1| c_{e_i}, \theta^{(j)}] \log(w_g^{(j)}  \mathcal{N}(c_{e_i}; \mu_g^{(j)}, \Sigma_g^{(j)})) \\
         &= \sum\limits_{i=1}^n \sum\limits_{g=1}^m \gamma_{ig}^{(j)} \log(w_g^{(j)}  \mathcal{N}(c_{e_i}; \mu_g^{(j)}, \Sigma_g^{(j)})) \\
    \end{aligned}
\end{equation}
where
\begin{equation}
     \gamma_{ig}^{(j)} \triangleq  \frac{w_g^{(j)} \mathcal{N}(c_{e_i}; \mu_g^{(j)}, \Sigma_g^{(j)})}{\sum\limits_{g^{\prime}=1}^m w^{(j)}_{g^{\prime}} \mathcal{N}(c_e; \mu_{g^{\prime}}^{(j)}, \Sigma_{g^{\prime}}^{(j)})}.
\end{equation}

Using this information, the updated $\theta$ estimates are calculated in the Maximization step.

\subsubsection{Maximization step}
The new parameter update $\theta^{(j+1)}$ is obtained by maximizing $Q(\theta|\theta^{(j)})$ with respect to $\theta$ as shown below,
\begin{equation}
    \begin{aligned}
     \theta^{(j+1)} = \arg \max_{\substack{\theta}}  Q(\theta|\theta^{(j)}).
    \end{aligned}
\end{equation}

The updated GMM parameters using this procedure can be computed as follows. The updated weight is given by
\begin{equation}
    \begin{aligned}
       w_g^{(j+1)} = \frac{\sum\limits_{g=1}^m \gamma_{ig}^{(j)}}{n}.
    \end{aligned}
\end{equation}

The updated mean is given by 
\begin{equation}
    \begin{aligned}
       \mu_g^{(j+1)} = \frac{\sum\limits_{g=1}^m \gamma_{ig}^{(j)} c_{e_i}}{\sum\limits_{g=1}^m \gamma_{ig}^{(j)}} .
    \end{aligned}
\end{equation}

The updated covariance matrix is given by
\begin{equation}
    \begin{aligned}
       \Sigma_g^{j+1} = \frac{1}{\sum\limits_{g=1}^m \gamma_{ig}^{(j)}}  \sum\limits_{i=1}^n \gamma_{ig}^{(j)} (c_{e_i} - \mu_g^{(j)}) (c_{e_i} - \mu_g^{(j)})^T.
    \end{aligned}
\end{equation}

After EM converges,
the cluster membership, $z$, given by $z_i =  \arg \max_{\substack{g}} \:\: P( z_i = \zeta_g| c_{e_i}, \theta)$, is
used to cluster the variables into different Gaussian components.

\subsubsection{Clustering step}

Let $\rho_g$ denote the set of indices of data points of $c_e$ that are classified into the $g^{th}$ Gaussian component and let $n_g$ denote the cardinality of this set. Then, the data points of $c_e$ which belong to the $g^{th}$ Gaussian component can be denoted by $c_{e_g} = \{c_{e_{g_i}}\}_{i \in \rho_g}, $  \textcolor{black}{$c_{e_g} \in \mathbb{R}^{n_g  \times 1}$}. 
As the parameter estimation is carried out using noisy measurements, 
both $c$ and $D$ are also clustered into $m$ different bins.
This results in $c_{g} = \{c_{{g_i}}\}_{i \in \rho_g}, \textcolor{black}{c_g \in \mathbb{R}^{n_g  \times 1}}$ and $D_{g} = \{D_{{g_i}}\}_{i \in \rho_g}, \textcolor{black}{D_g \in \mathbb{R}^{n_g  \times p}}$. 
After clustering the data points into different Gaussian components,
the following logic is used to 
set-up the parameter estimation problem.
 Substituting $z_g^i$ 
 into \eqref{Complete log-likelihood}, a modified log-likelihood function can be computed for the GMM as 
 
\begin{equation}                                             
\label{Modified_log_L_GMM}
    \begin{aligned}
         \log ( L(c_e; \mu, \Sigma))   &= \sum\limits_{g=1}^m \sum\limits_{i=1}^n  z_g^i \log(w_g \mathcal{N}( c_{e_i}; \mu_g, \Sigma_g)) \\
         &= \sum\limits_{g=1}^m \sum_{i\in \rho_g} \log(w_g \mathcal{N}( c_{e_i}; \mu_g, \Sigma_g)).\\
    \end{aligned}
\end{equation}

In turn, \eqref{Modified_log_L_GMM} can be expressed as
\begin{equation}                                             
\label{GMM_MLE_Min_Equivalence}
    \begin{aligned}
         \log ( L(c_e; \mu, \Sigma))  
           &= K -  \sum\limits_{g=1}^m  (c_{e_g} - \mu_g)^T \Sigma_g^{-1} (c_{e_g}-\mu_g),
    \end{aligned}
\end{equation}
where, $K = \sum\limits_{g=1}^m \sum\limits_{i\in \rho_g}  \log(w_g)  -\frac{d}{2}\log(2 \pi) -\frac{1}{2}\log (|\Sigma_g|)$. The variable $\Sigma_g$ here denotes the covariance matrix of $g^{th}$ Gaussian component of 
$c_e$, and the variable $\mu_g$ denotes the mean vector of $g^{th}$ Gaussian component of 
$c_e$. For optimal parameter estimation, the log-likelihood has to be maximized with respect to the parameters. It can be observed that maximizing the log-likelihood in \eqref{GMM_MLE_Min_Equivalence} is equivalent 
to minimizing $\sum\limits_{g=1}^m  (c_{e_g} - \mu_g)^T \Sigma_g^{-1} (c_{e_g}-\mu_g)$.
This implies that the optimal parameter estimate under GMM noise environment can be obtained when the \textit{sum of squares of standardized errors} (SSSE) is minimized, where the standardization is done with respect to 
the individual Gaussian component to which the noise data sample belongs. 
This becomes the functional basis for the optimization problems for the parameter estimation step as explained in the next two sub-sections.

\subsection{Parameter Estimation when Non-Gaussian Noise is only Present in Dependent Variables}
\label{ParEstDep}

For the scenario in which non-Gaussian noise (expressed as a GMM) is present only in the dependent variables, the parameter estimation can be formulated as a minimization problem whose objective is to minimize the SSSE of the noise variable and the constraints are the error definitions.
This is mathematically described by 
\begin{equation}
\label{Level 1 - Minimization}
\begin{aligned}
    \phi = \arg \min_{\substack{x, c_e }} \:\: & \sum\limits_{g=1}^{m} \frac{1}{2} [\Sigma_g^{-\frac{1}{2}} (c_{e_{g}} - \mu_g)]^T [\Sigma_g^{-\frac{1}{2}} (c_{e_{g}} - \mu_g)] \\
    \text{such that } &[c_{g} - D_{g} x  - c_{e_{g}}]=0 \:\: \forall g \in [m]. \\
\end{aligned}
\end{equation}

Using a vector of Lagrange multipliers, \textcolor{black}{$\lambda_g \in \mathbb{R}^{n_g  \times 1}$}, for each Gaussian component, \eqref{Level 1 - Minimization} is expressed as an unconstrained optimization problem as shown below
\begin{equation}
\label{Level 1 - unconstrained Minimization}
\begin{aligned}
    \phi = \arg \min_{\substack{x, c_e }} \:\: & \sum\limits_{g=1}^{m} \frac{1}{2}  [\Sigma_g^{-\frac{1}{2}} (c_{e_{g}} - \mu_g)]^T [\Sigma_g^{-\frac{1}{2}} (c_{e_{g}} - \mu_g)] \\
    &+ \sum\limits_{g=1}^{m} [c_{g} - D_{g} x - c_{e_{g}}]^T \lambda_g. \\
\end{aligned}
\end{equation}

The optimal parameter estimate is found by computing the $\hat{x}$ that minimizes \eqref{Level 1 - unconstrained Minimization}. This is done by simultaneously solving $\pdv{\phi}{c_{e_{g}}} = 0,  g \in [m]$, $\pdv{\phi}{\lambda_g}=0,  g \in [m]$, and $\pdv{\phi}{x}=0$, to yield the following 
equation for the parameter estimate:

\begin{equation}
\label{Level1_Closed_form_Expression}
\begin{aligned}
\hat{x} &=  \left(\sum\limits_{g=1}^{m}  \Sigma_g^{-1}  D_{g}^T D_{g}\right)^{-1} \left(\sum\limits_{g=1}^{m}  \Sigma_g^{-1} D_{g}^T ( c_{g} - \mu_g) \right).
\end{aligned}
\end{equation}

Equation \eqref{Level1_Closed_form_Expression} can be directly used to compute the optimal parameter estimate when the noise characteristics and cluster membership of data points of the GMM is available. In the actual setting where the measurement noise characteristics are unknown, the noise estimation procedure explained in Section \ref{NoiEst} must be used in conjunction with this parameter estimation step and iterated until convergence, leading to a joint estimation of noise and parameters.
\textcolor{black}{However, Section \ref{NoiEst} assumes prior knowledge of $m$.
For instance, Ahmad et al. have shown in \cite{ahmad2019statistical} that PMU measurement noise (particularly the noise in the current phasors) can be represented using a three component GMM.
However, as the power system is non-stationary, there is no guarantee that $m$ will always be equal to three.  
A strategy to circumvent the need for the \textit{a priori} knowledge of $m$ is proposed below.}

\textcolor{black}{In absence of the prior knowledge of $m$, the proposed methodology for joint estimation of noise and parameters must be repeated for a range of values of $m$, say, $m=1$ to $m = m_{max}$. 
A reasonable value of $m_{max}$ could be ten.
Then, an information theory-based model selection criterion called the Bayesian information criterion (BIC) can be used to identify the most appropriate value for $m$ \cite{neath2012bayesian}. The BIC has two components - one component rewards the goodness of fit of the model, while the other component penalizes the model complexity.
Thus, the optimal number of GMM components will be the $m$ for which the lowest value of BIC is obtained.
Note that by removing the need for the \textit{a priori} knowledge of $m$, this strategy further improves the robustness of the proposed approach.
The flowchart describing the overall procedure is presented in Fig. \ref{Fig_Joint_PE_NE_Flowchart}.}

In Fig. \ref{Fig_Joint_PE_NE_Flowchart},
\textcolor{black}{for each value of $m$,}
the expectation and maximization steps learn the characteristics 
of the 
GMM that approximates the noise vector.
The cluster membership is used to cluster the data points in the noise vector
into $m$ 
Gaussian components. 
These, then become inputs to
the parameter estimation step, which is carried out using  \eqref{Level1_Closed_form_Expression}. 
\textcolor{black}{For 
each value of $m$,
noise and parameters estimates in presence of non-Gaussian noise in the dependent variables are obtained in an iterative manner.
The optimal estimates of noise and parameters correspond to the value of $m$ that gives the lowest value of BIC. This concludes the flowchart.}





\subsection{Parameter Estimation when Non-Gaussian Noise is Present in Dependent and Independent Variables}
\label{ParEstInd}

The second scenario accounts for measurement noise in both the dependent as well as the independent variables, and are known as EIV problems \cite{schennach2016recent}. 
The measurement model for this scenario is mathematically described by \eqref{linear_system_relationship} and \eqref{EIV measurement model}.
The noise estimation methodology described in Section \ref{NoiEst} is leveraged here to solve the EIV problem in presence of non-Gaussian noise in both dependent and independent variables. 
The $g^{th}$ Gaussian component of the standardized dependent variable noise vector is denoted as
 \begin{equation}
     c_{e_{g_N}} = [\Sigma_{c_g}^{-\frac{1}{2}} \: (c_{e_{g}} - \mu_{c_g})].
 \end{equation}

Similarly, the $g^{th}$ Gaussian component of the standardized independent variable noise vector can be denoted as
 \begin{equation}
     D_{e_{g_N}} = [\Sigma_{D_g}^{-\frac{1}{2}} \: (D_{e_{g}} - \mu_{D_g})].
 \end{equation}

Note that $c_{e_{g_N}}^T  c_{e_{g_N}}$ denotes the SSSE in $c_{e_{g}}$. 
However, as $D_{e_{g_N}}$ is a matrix, it has to be first converted to a column vector form before it can be converted to an SSSE form. Let $\vectext(D_{e_{g_N}})$ denote the vectorization operation of the matrix $D_{e_{g_N}}$, where the matrix is converted to a column vector by stacking the columns of the matrix on top of each other. Then, using the properties of the Kronecker product, $\otimes$, the 
resulting minimization problem for optimal parameter estimation for the EIV problem can be mathematically described by:

\begin{equation}
\label{OpParaEIV}
\begin{aligned}
    \phi &= \arg \min_{\substack{x }} \:\:  \sum_{g=1}^{m} \frac{1}{2} ( c_{e_{g_N}})^T ( c_{e_{g_N}}) \\
      &+ \sum_{j=1}^{m} \frac{1}{2} (\vectext( D_{e_{g_N}}))^T  (\vectext( D_{e_{g_N}}))\\
    \text{s.t. } & [c_{g} - (x^T \otimes I_{n_g}) ( \vectext(D_{g})- \vectext(D_{e_{g}}) ) - c_{e_{g}} ]\\
    &\:\: \forall g \in [1, \dots, m].
\end{aligned}
\end{equation}

Using the method of Lagrange multipliers, \eqref{OpParaEIV} can be written as an unconstrained objective function as

\begin{equation}
\begin{aligned}
    \phi &= \arg \min_{\substack{x }} \:\:  \sum_{g=1}^{m} \frac{1}{2} ( c_{e_{g_N}})^T ( c_{e_{g_N}}) \\
      &+ \sum_{j=1}^{m} \frac{1}{2} (\vectext( D_{e_{g_N}}))^T  (\vectext( D_{e_{g_N}}))\\
  &+ \sum_{g=1}^{m} [c_{g} - (x^T \otimes I_{n_g}) ( \vectext(D_{g}) - \vectext(D_{e_{g}})) - c_{e_{g}} ]^T \lambda_g\\
\end{aligned}
\end{equation}


\begin{figure}[t]
\centering
\includegraphics[width=0.48\textwidth]{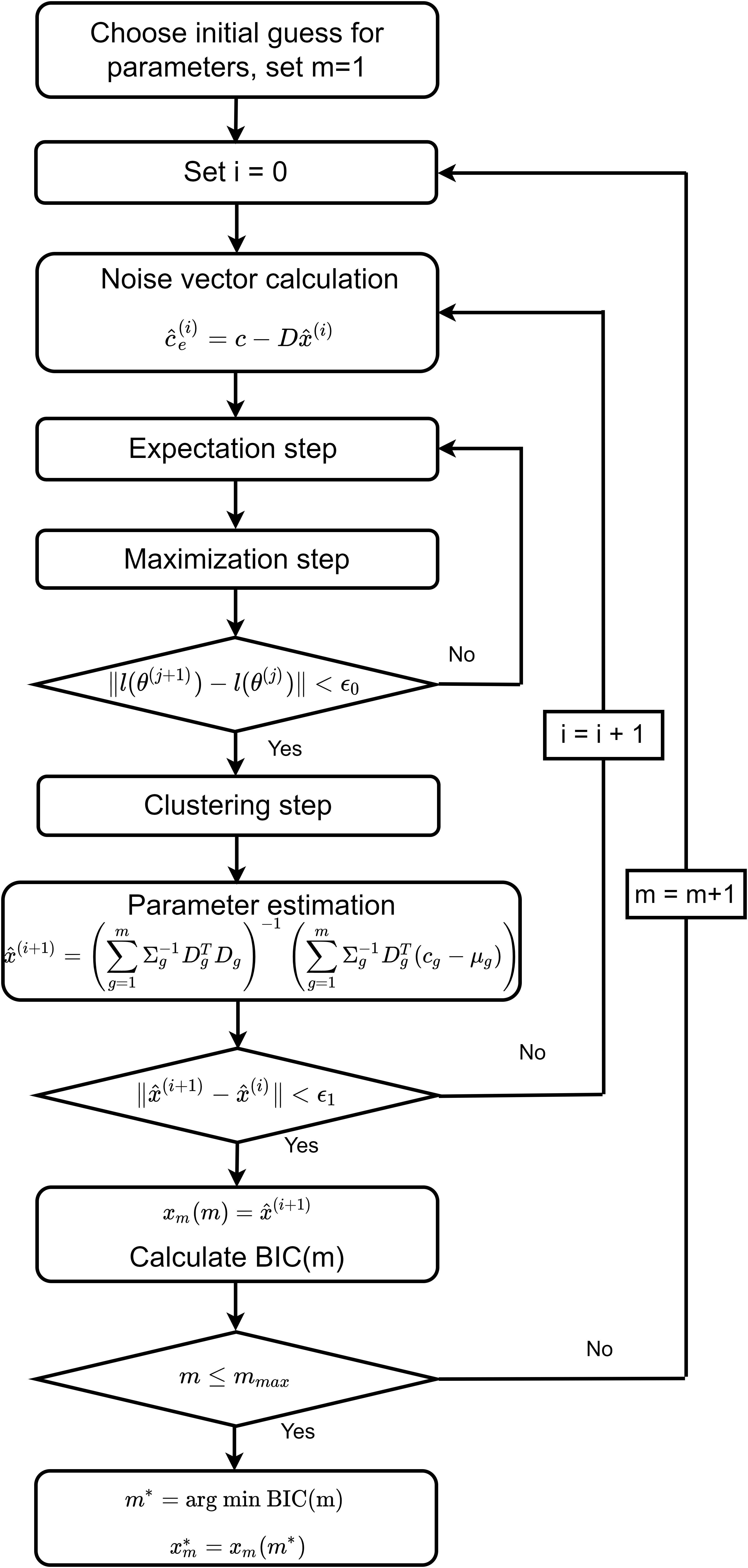}
\caption{\textcolor{black}{Iterative method for parameter estimation in presence of non-Gaussian noise in dependent variable}}
\label{Fig_Joint_PE_NE_Flowchart}
\end{figure}

The optimal parameter estimate for this objective function is the solution to the following non-linear system of equations:
\begin{equation}
\label{Gen_EIV_Solution1}
    \begin{aligned}
      f(x) = \sum_{g=1}^{m} (D_{g} - D_{e_{g}})^T \lambda_g  &= 0 .\\
    \end{aligned}
\end{equation}
where,

\begin{subequations}
\label{Gen_EIV_Solution_sub_variables}
    \begin{align}
        \lambda_g  &= (\Sigma_{ net g})^{-1} ( c_{g} - D_{g} x  - \mu_{ net g})\\
       \mu _{net g} &= \mu_{c_g} - \sum\limits_{j=1}^p \mu_{D_g} \times {x}_j \\
        \Sigma_{ net g} &= \Sigma_{c_g} + \sum\limits_{j=1}^p \Sigma_{D_g} \times {x}_j^2.
    \end{align}
\end{subequations}


In \eqref{Gen_EIV_Solution_sub_variables}, $j$ denotes the $j^{th}$ column for D and $j^{th}$ parameter for $x$. The optimal parameters are found by solving 
\eqref{Gen_EIV_Solution1} using Newton's method:

\begin{equation}
       \label{NM_Updation_1st_deri}
           \begin{aligned}    
          \begin{bmatrix}
             x^{(k+1)}      
          \end{bmatrix} = \begin{bmatrix}
             x^{(k)}      
          \end{bmatrix} - \begin{bmatrix}
            \text{Jac}(f(x^{(k)}))      
          \end{bmatrix}^{-1}   \begin{bmatrix}
             f(x^{(k)})     
          \end{bmatrix}
        \end{aligned}
\end{equation} 
where, $\text{Jac}(f(x^{(k)}))$ denotes the Jacobian of $f(x^{(k)})$.
The components of the optimal noise estimates of $c_e$ and $D_e$ can be obtained from the following equation,

 

\begin{subequations}
\label{EIV-noise-estimates}
    \begin{align}
        \hat{c}_{e_{g}} &= \Sigma_{c_g} \lambda_g + \mu_{c_g}\\
      \hat{D_e}_{j_{g}}&= -x_j \Sigma_{D_{g}} \lambda_g +  \mu_{D_{g}}.
    \end{align}
\end{subequations}


The GMM noise characteristics and cluster membership can be obtained by applying EM to the noise estimates.
Since the parameter estimation step uses the results from the noise estimation step and vice-versa, 
they are repeated 
until convergence. 
\textcolor{black}{
Lastly, the
noise and parameter estimation is performed iteratively for $ \forall m \in [1,m_{max}]$, and BIC is used to select the optimal $m$, and noise and parameter estimates.
The overall process is described as an \textbf{Algorithm}\ref{Algorithm-PE-EIV-GMM2} below. 
As this algorithm solves an \textbf{E}IV problem using a combination of \textbf{G}MMs, \textbf{L}agrange multipliers, and \textbf{E}M, it is henceforth called the EGLE Algorithm. 
}



\begin{algorithm}
    \renewcommand{\thealgorithm}{} 
    \caption{\textcolor{black}{EGLE Algorithm for Robust Parameter Estimation for EIV Problem with Non-Gaussian Noise}}
    \label{Algorithm-PE-EIV-GMM2}
    \hspace*{\algorithmicindent} \text{Input: Noisy Measurements (c, D)}, $x^{(0)}, \epsilon_0, \epsilon_1, \epsilon_2, m_{max}$ \\
    \hspace*{\algorithmicindent} \text{Output: Optimal Parameter ($\hat{x}$), Noise Estimate ($\hat{c_e}$, $\hat{D_e}$)} 
    \begin{algorithmic}[1]
    \Procedure{}{}
    \For{$m = 1$ to $m_{max}$}
    \State Initialize $x, {c_e}, {D_e}$
    \State $  \mu_c, \Sigma_c, w_c \gets \text{EM}(\hat{c_e})$ (See Section \ref{NoiEst})
    \State $ \mu_D, \Sigma_D, w_D \gets \text{EM}(\hat{D_e}) $ (See Section \ref{NoiEst}) 
    \For{$i = 1$ to $i_{max}$}
            \State Compute  $ \hat{c}_{e_{g}}$ \textcolor{black}{using (\ref{EIV-noise-estimates}a)}
            \State Compute $\hat{D_e}_{j_{g}}$ \textcolor{black}{using (\ref{EIV-noise-estimates}b)}
            \State $\hat{c_e} \gets \text{Concatenate}(\hat{c}_{e_{g}}),\: \forall g \in [m]$
            \State $\hat{D_e} \gets \text{Concatenate}(\hat{D}_{e_{g}}),\: \forall g \in [m]$
            \State $  \mu_c, \Sigma_c, w_c \gets \text{EM}(\hat{c_e})$ (See Section \ref{NoiEst})
            \State $ \mu_D, \Sigma_D, w_D \gets \text{EM}(\hat{D_e}) $ (See Section \ref{NoiEst})
            \For{$k = 1$ to $k_{max}$}
            \State Compute $ \mu_{ net g}$ \textcolor{black}{using (\ref{Gen_EIV_Solution_sub_variables}a)}
            \State Compute $ \Sigma_{ net g}$ \textcolor{black}{using (\ref{Gen_EIV_Solution_sub_variables}b)}
            \State Compute $ \lambda_{ net g}$ \textcolor{black}{using  (\ref{Gen_EIV_Solution_sub_variables}c)}
            \State Compute $f(x)$ \textcolor{black}{using  \eqref{Gen_EIV_Solution1}}
             \State Find $x^{(k+1)}$ \textcolor{black}{using \eqref{NM_Updation_1st_deri}}
                    \If {$\| x^{(k+1)} - x^{(k)} \|_2 < 
                    \epsilon_2$}
                    \State \textbf{break}.
                    \EndIf
                    \State $k \gets k+1$.
            \EndFor
    \State $x^{(i+1)} = x^{(k+1)} $
     \If {$\| x^{(i+1)} - x^{(i)} \|_2 < 
    \epsilon_1$}
    \State $x_{m} (m) = x^{(i+1)}$
                    \State \textbf{break.} 
    \EndIf
    \State $i \gets i+1$.
    \EndFor
    \State Calculate $\mathrm{BIC}(m)$
     \State $m \gets m+1$.
    \EndFor
    \State $m^{*} = \text{arg min} (\mathrm{BIC}(m))$
    \State $x^{*} = x_m(m^*)$
    \EndProcedure
    \end{algorithmic}
    \end{algorithm}
      \vspace{-5mm}
\vspace{4 mm} \textcolor{black}{The inputs for the \textbf{EGLE Algorithm}\ref{Algorithm-PE-EIV-GMM2} are the noisy measurements, an initial guess for the parameters, the tolerances ($\epsilon_0, \epsilon_1, \epsilon_2$), and $m_{max}$. The tolerance values can be chosen based on the desired estimation accuracy and speed requirements. 
The outermost loop finds the optimal number of Gaussian components ($m^*$).
The loop in the middle performs noise estimation using EM (see Section \ref{NoiEst}) and provides inputs to the innermost loop. 
The innermost loop performs parameter estimation by solving \eqref{Gen_EIV_Solution1} using Newton's method.
The parameter estimate obtained from the innermost loop serves as an input to the loop in the middle.
The tolerance $\epsilon_2$ determines the termination of the parameter estimation loop, whereas $\epsilon_1$ determines when the joint noise and parameter estimation is completed for each value of $m$.
The value of BIC corresponding to every value of $m$ is also calculated.
Finally, the optimal value of $m$ is found using the index of the minimum value of $\mathrm{BIC}(m)$. The $m^*$ thus obtained gives the optimal number of GMM components that must be used to approximate the non-Gaussian noise, while the parameter estimate corresponding to that $m^*$ (namely, $x_m(m^*)$) is the optimal parameter estimate, $x^{*}$.
}

\textcolor{black}{Note that when $D$ is noise-free, then the \textbf{EGLE Algorithm}\ref{Algorithm-PE-EIV-GMM2} simplifies to the flowchart shown in Fig. \ref{Fig_Joint_PE_NE_Flowchart}. As such, EGLE can be used when non-Gaussian noise is only present in the dependent variables as well as when non-Gaussian noise is present in both the dependent and the independent variables.  
The results obtained using EGLE for simulated and actual PMU data are described in the following sections.}

\section{Results: TLPE using Simulated PMU Data}
\label{Results}
\textcolor{black}{The IEEE 118-bus system was used as the test system for the analysis conducted in this section.}
The system was solved using MATPOWER \cite{zimmerman2010matpower}, an open-source package in MATLAB, to generate the true (noise-free) voltage and current phasor measurements.
The system loading was varied by 40\% over multiple time instants to create linearly independent sets of measurements; note that this type of variation naturally occurs in a power system during the morning load pick-up \cite{gao2012dynamic}.
The noisy measurements were generated by adding a suitable noise (depending on the scenario considered) to the true values.
The proposed algorithm requires an initial guess for $x$, $c_e$, and $D_e$.
It has been observed in \cite{kusic2004measurement, mansani2018estimation} that although the line parameters vary during regular operation, they lie within 30\% of their values mentioned in a power utility's database.
Therefore, the values of the line parameters in a utility's database are suitable initial values of $x$.
A zero-mean Gaussian distribution was assumed as the initial guess for $c_e$ and $D_e$.

For comparing the performance of the EGLE with other methods, the absolute relative error ($\mathrm{ARE}$) index was used. It is mathematically defined as
\begin{equation}
    ARE = \frac{|x_{est} - x_{true}|}{x_{true}},
\end{equation}
where, $x_{est}$ is the estimated parameter, and $x_{true}$ is the true parameter.
For comparing the performance for a given parameter over many Monte Carlo (MC) runs, the mean value of ARE, ($\mathrm{MARE}$), and the standard deviation of the ARE, ($\mathrm{SDARE}$), were computed.
Similarly, to compare performance across all the parameters, the mean value of the net $\mathrm{ARE}$ ($\mathrm{MARE_{net}}$) and its standard deviation ($\mathrm{SDARE_{net}}$) were computed.
All the analyses performed in this paper were conducted using Python programming language.

\subsection{Non-Gaussian Noise only in Dependent Variables}
\label{Results-GMM Noise in Dependent variables}
\begin{figure}
        \centering
        \begin{subfigure}[b]{0.425\textwidth}
            \centering
            \includegraphics[width=\textwidth]{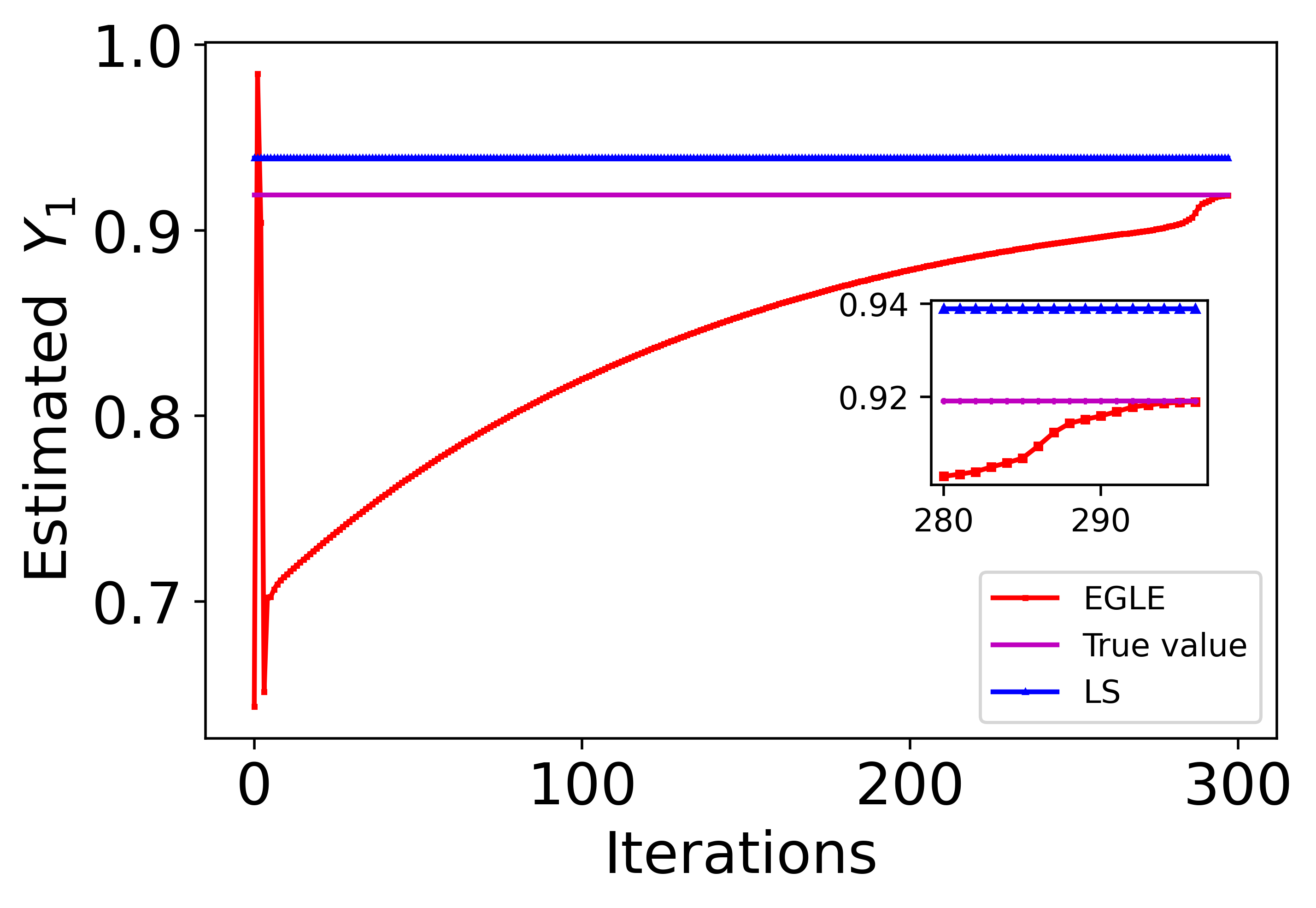}
            \caption[Real(Vdiff)]%
            {{\small $Y_1$ vs Iterations}}    
            \label{TLPE_PS_Convergence_Level1_Y1}
        \end{subfigure}
        \hfill
        \begin{subfigure}[b]{0.425\textwidth}  
            \centering 
            \includegraphics[width=\textwidth]{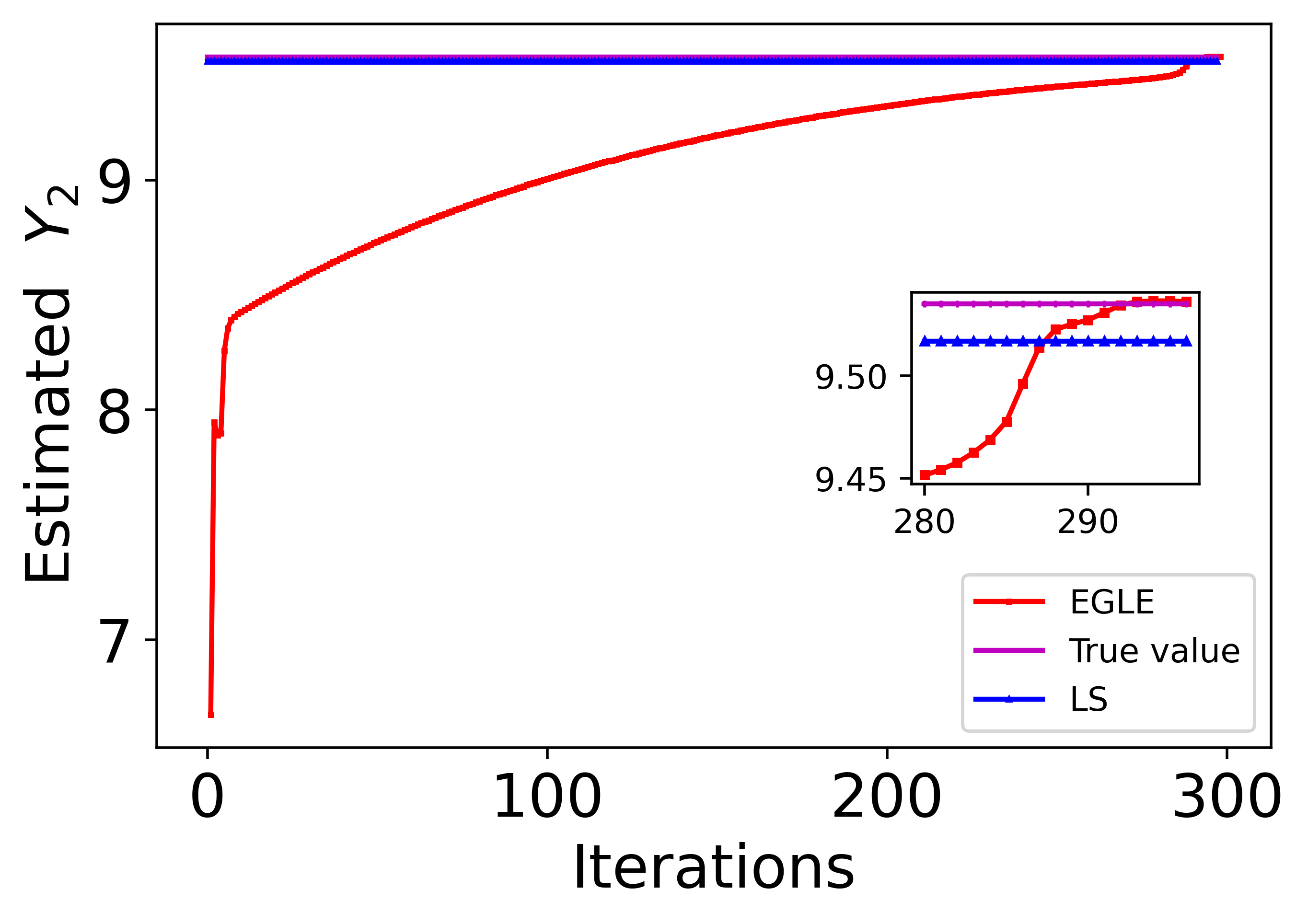}
            \caption[]%
            {{\small  $Y_2$  vs Iterations }}    
            \label{TLPE_PS_Convergence_Level1_Y2}
        \end{subfigure}
      \label{TLPE_PS_Convergence_Level1b}
        \centering
        \begin{subfigure}[b]{0.425\textwidth}
            \centering
            \includegraphics[width=\textwidth]{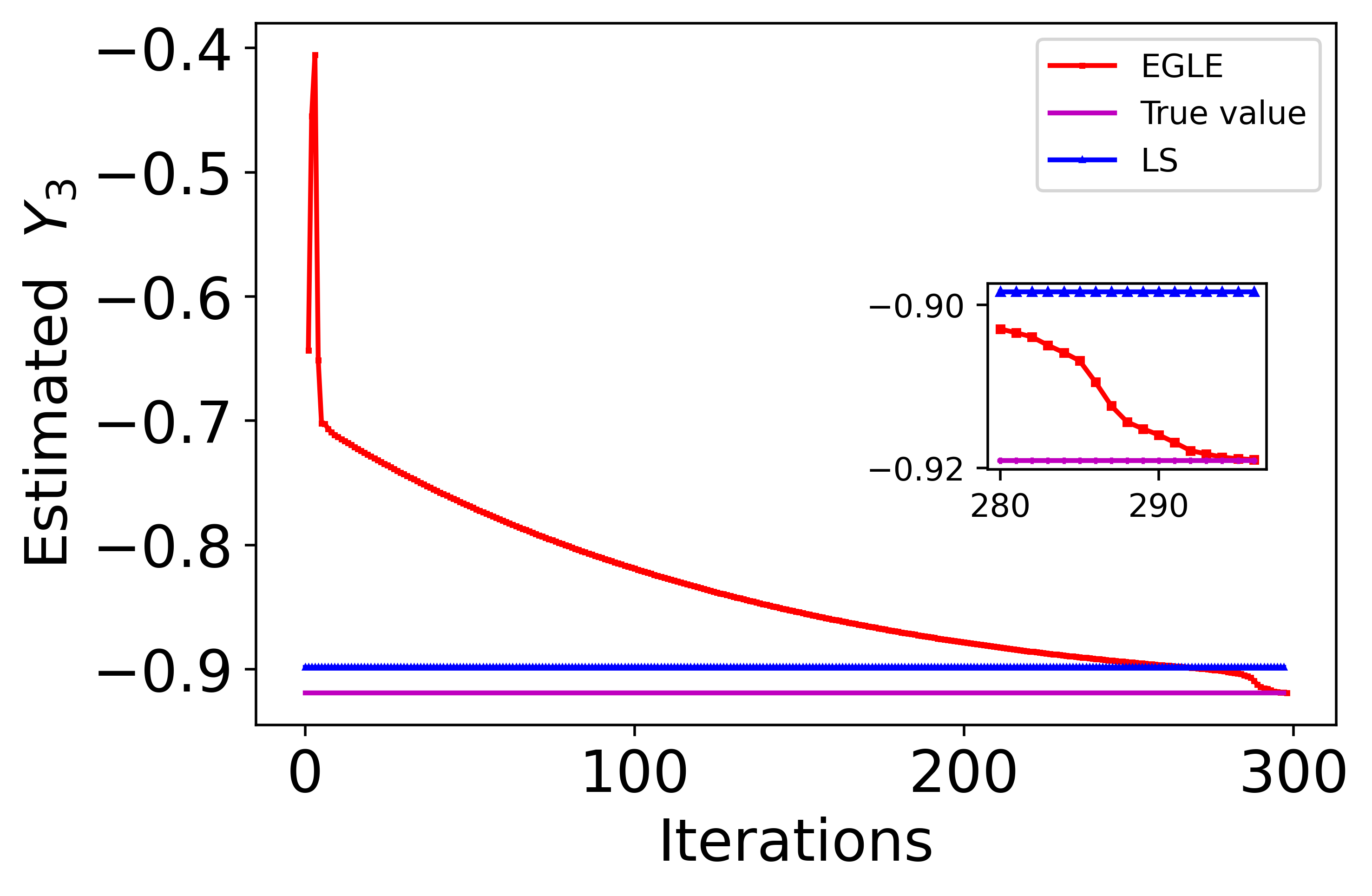}
            \caption[Real(Vdiff)]%
            {{\small $Y_3$ vs Iterations}}    
            \label{TLPE_PS_Convergence_Level1_Y3}
        \end{subfigure}
        \hfill
        \begin{subfigure}[b]{0.425\textwidth}  
            \centering 
            \includegraphics[width=\textwidth]{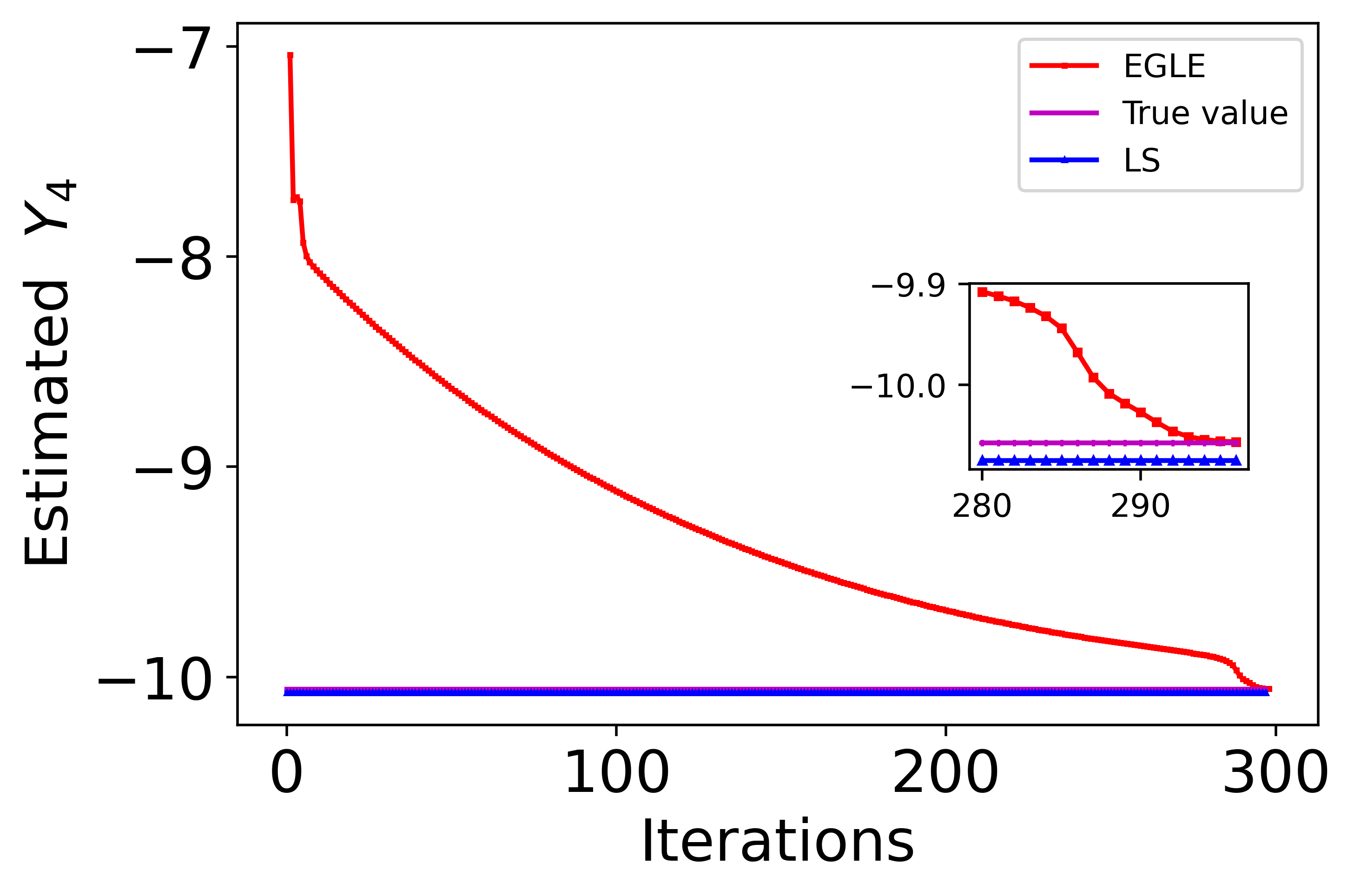}
            \caption[]%
            {{\small  $Y_4$  vs Iterations }}    
            \label{TLPE_PS_Convergence_Level1_Y4}
        \end{subfigure}
        \caption{\textcolor{black}{Convergence of the proposed method with GMM noise in dependent variables for $L_{38-65}$ of the IEEE 118-bus system}}
    \end{figure}

In this sub-section, a two component GMM measurement noise of mean ($\begin{bmatrix}
   0 & 0.005
\end{bmatrix}$), standard deviation  ($\begin{bmatrix}
   0.0015 & 0.0015
\end{bmatrix}$), and weight  ($ \begin{bmatrix}
   0.3 & 0.7
\end{bmatrix}$) was added to the currents to create the noisy dependent variables, while the independent variables (voltages) were kept noise-free.
The convergence of EGLE for 
the line which connects buses 38 and 65 ($L_{38-65}$) of the IEEE 118-bus system is shown in Fig. \ref{TLPE_PS_Convergence_Level1_Y1}-\ref{TLPE_PS_Convergence_Level1_Y4}.
The figures also compare the performance of the proposed method with LS for all four $Y$ parameters.
The colors magenta, red, and blue correspond to the true value, and estimates obtained from EGLE and LS, respectively. 
It can be observed from the figures that for each of the parameters, the estimate obtained using EGLE is: (a) very close to the true value, and (b)
better than the LS estimate. It is also obvious that the superior performance in estimating the $Y$ parameters will lead to better performance in the estimation of the actual line parameters from the $Y$ values.



\subsection{Non-Gaussian Noise in Dependent and Independent Variables}
\label{Results-GMM Noise in Dependent and independent variables}
\begin{figure}[ht]
            \centering
            \includegraphics[width=0.40\textwidth]{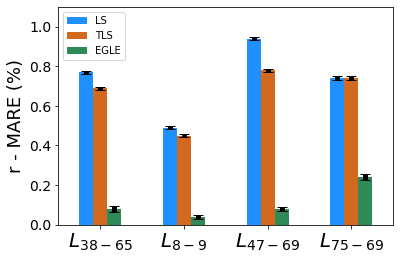}
            \caption{$\mathrm{MARE}$ of the estimated resistance for four lines of the IEEE 118-bus system}
            \label{Different_Lines_Resistance_Estimates}
\end{figure}

\begin{figure}[ht]
            \centering
            \includegraphics[width=0.40\textwidth]{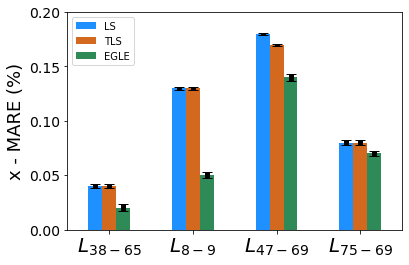}
            \caption{$\mathrm{MARE}$ of the estimated reactance for four lines of the IEEE 118-bus system}
            \label{Different_Lines_Reactance_Estimates}
\end{figure}

\begin{figure}[ht]
            \centering
            \includegraphics[width=0.40\textwidth]{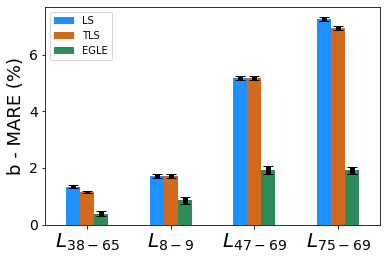}
            \caption{$\mathrm{MARE}$ of the estimated susceptance for four lines of the IEEE 118-bus system}
            \label{Different_Lines_Susceptance_Estimates}
\end{figure}

This sub-section corresponds to the most general setting in which both dependent and independent variables have non-Gaussian noise.
The two component GMM noise that was defined in Section \ref{Results-GMM Noise in Dependent variables} was also used here, except that this time it was added to both current and voltage measurements. 
The results corresponding to two 345 kV lines and two 138 kV lines of the IEEE 118-bus system are analyzed below. The 345 kV lines are between buses 38 and 65, and between buses 8 and 9. The 138 kV lines are between buses 47 and 69, and between buses 75 and 69. 
The comparison of the resistance, reactance, and susceptance estimates obtained using LS, TLS, and EGLE for these four lines is shown in Fig. \ref{Different_Lines_Resistance_Estimates}, Fig.  \ref{Different_Lines_Reactance_Estimates}, and Fig.  \ref{Different_Lines_Susceptance_Estimates}, respectively. 


To draw reliable statistical inferences, the experiment was repeated 1,000 times by randomly generating noise based on the specified GMM characteristics and a random initial guess for the parameter estimates that was within $\pm 30\%$ of the true value. 
The $\mathrm{MARE}$ for LS, TLS, and EGLE 
for all the MC runs 
are shown as bar plots in \color{black}Fig. \ref{Different_Lines_Resistance_Estimates}-\ref{Different_Lines_Susceptance_Estimates}\color{black}.  The $\mathrm{SDARE}$ 
are also calculated, a low value of which is an indication of the consistency of the results. The $\mathrm{SDARE}$ is displayed as black colored error bars in \color{black}Fig. \ref{Different_Lines_Resistance_Estimates}-\ref{Different_Lines_Susceptance_Estimates}\color{black}.

From \color{black}Fig. \ref{Different_Lines_Resistance_Estimates}-\ref{Different_Lines_Susceptance_Estimates}\color{black},  it can be observed that EGLE consistently performs better than LS and TLS when estimating line
parameters in presence of GMM noise. 
This also translates to superior performance of EGLE in terms of the $\mathrm{MARE_{net}}$. 
For example, for $L_{38-65}$,
the $\mathrm{MARE_{net}}$ for LS and TLS methods were $1.55 \%$ and $1.33 \%$, respectively, whereas the proposed method had a $\mathrm{MARE_{net}}$ of only $0.40 \%$. The $\mathrm{SDARE_{net}}$ for the three methods were 0.06\%, 0.06\%, and 0.10\% respectively, indicating consistency of the estimates.  A similar trend was found for the other lines as well. 
\nolinebreak
\nolinebreak
\nolinebreak
\textcolor{black}{It was also observed that the relative accuracies of the parameter estimates for the three algorithms across all four lines was in accordance with the conditioning 
of the $D^T  D$ matrix.}

\subsection{\textcolor{black}{Comparison of EGLE with a Denoising-followed-by-(conventional)-estimation Approach}}
\label{Comparison with signal denoising based approach}
\textcolor{black}{The results obtained in the previous two sub-sections demonstrate the superior performance of EGLE over the conventional approaches in presence of non-Gaussian noise in the measurements. 
However, as TLPE is an \textit{offline} process, an alternate strategy could be to first remove the non-Gaussian noise from the measurements, and then perform estimation using conventional approaches.
This strategy, called the denoising-followed-by-(conventional)-estimation approach, is compared with EGLE in this sub-section.}

\begin{figure}[ht]
        \centering
            \begin{subfigure}[b]{0.40\textwidth}  
            \centering 
            \includegraphics[width=\textwidth]{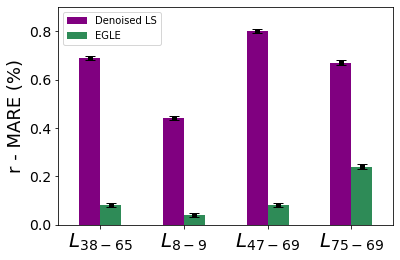}
            \caption[]%
            {{\small Resistance estimates}}    
            \label{denoise_r}
        \end{subfigure}
        \hfill
        \begin{subfigure}[b]{0.40\textwidth}
            \centering
            \includegraphics[width=\textwidth]{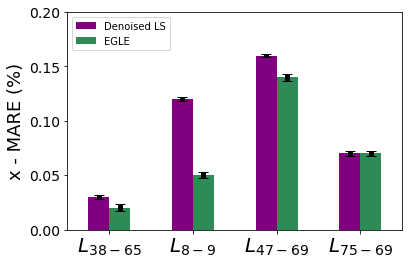}
            \caption[Real(Vdiff)]%
            {{\small Reactance estimates}}    
            \label{denoised_x}
        \end{subfigure}
         \hfill
        \begin{subfigure}[b]{0.40\textwidth}  
            \centering 
            \includegraphics[width=\textwidth]{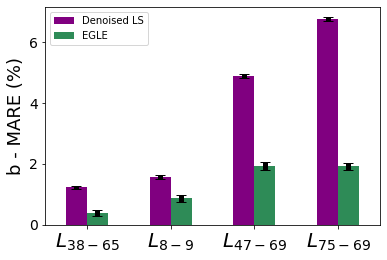}
            \caption[]%
            {{\small Susceptance estimates}}    
            \label{denoised_b}
        \end{subfigure}
         \caption{\textcolor{black}{Performance comparison of EGLE and a denoising-followed-by-(conventional)-estimation approach for TLPE of four lines of the IEEE 118-bus system}}
            \label{denoised_rxb}
\end{figure}

\textcolor{black}{The noisy data created in Section \ref{Results-GMM Noise in Dependent and independent variables} was used for this analysis.
The denoising was done for both voltage and current phasor measurements using a moving window median absolute deviation filter that was developed in \cite{ahmad2019statistical} to detect non-Gaussian noise in PMU measurements. Subsequently, the LS method was employed for TLPE. The results that were obtained when this approach was compared with EGLE are shown in Fig. \ref{denoised_rxb}.
In Fig. \ref{denoised_rxb}, the purple bars indicate the $\mathrm{MARE}$ obtained using the denoising approach whereas the green bars show the $\mathrm{MARE}$ obtained using EGLE. It is clear from the figure that EGLE performs better than the denoising based approach, particularly for the resistance and susceptance estimates.
One reason for this observation could be the sensitivity of the filter to the size of the window that was considered for analysis (the window size was set at 600 samples as mentioned in \cite{ahmad2019statistical}).
In summary, this analysis demonstrated the superiority of an integrated approach for noise and parameter estimation (EGLE) in comparison to a denoising-followed-by-(conventional)-estimation approach.
}

\subsection{\textcolor{black}{Comparison of EGLE with the Minimum Total Error Entropy (MTEE) Method}}
\label{Comparison with another iterative approach}
\textcolor{black}{The PMU-based TLPE problem is a linear regression problem in which the (line) parameters to be estimated change at a much slower rate than the speed at which the (PMU) measurements become available \cite{chatterjee2018error}. This, in turn, implies that TLPE is a \textit{static linear estimation} problem.
Recently, an information theoretic measure called \textit{total error entropy} has been proposed to perform parameter estimation for static linear estimation problems in which the dependent and independent variables have non-Gaussian noises \cite{shen2015minimum}. 
The resulting technique is called the MTEE method, and is also iterative; a brief overview of MTEE is provided in Appendix \ref{MTEE Overview}.
As both MTEE and EGLE solve the same type of estimation problems, in this sub-section, we perform a comparison of the two for the TLPE problem.
The noisy data created in Section \ref{Results-GMM Noise in Dependent and independent variables} was used for this analysis.
}

\textcolor{black}{To compare the performances of MTEE method with EGLE, two types of simulations were conducted. In the first set of simulations, the number of iterations were fixed at 200 and the estimation accuracy of the two methods was compared in terms of the $\mathrm{MARE}$ index. The initial guesses were kept the same for both the approaches. The results obtained are shown in Table \ref{MTEE-comparison}.
For ease of comparison, the results of LS and TLS are also provided.
It can be observed from the table that even though the MTEE method has lower $\mathrm{MARE}$ than the LS and TLS methods, 
it could not outperform EGLE. The trend was consistent across the resistance, reactance, and susceptance estimates as well as across all four transmission lines of the IEEE-118 bus system that were analyzed.
}

\begin{table}[H]
\color{black}
\centering
\caption{\textcolor{black}{Performance comparison of MTEE and EGLE (for fixed number of iterations)}}
\label{MTEE-comparison}
\begin{tabular}{|l|l|c|c|c|}
\hline
             &          & r (MARE (\%))   & x (MARE(\%))    & b (MARE(\%))   \\ \hline
    \multirow{4}{*}{$L_{38-65}$}   & LS    & 0.77 & 0.04  & 1.34 \\ \cline{2-5}
   & TLS    & 0.69 & 0.04  & 1.15 \\ \cline{2-5}
    & MTEE     & 0.53 & 0.03  & 0.98 \\ \cline{2-5}
 & EGLE & 0.09 & 0.02 & 0.39 \\ \hline
 \multirow{4}{*}{$L_{8-9}$ }  & LS    & 0.49 & 0.13  & 1.72 \\ \cline{2-5}   
 & TLS    & 0.45 & 0.13  & 1.71 \\ \cline{2-5}
  & MTEE     & 0.34 & 0.12  & 1.40 \\ \cline{2-5}
  & EGLE & 0.06 & 0.05 & 0.92 \\ \hline
  \multirow{4}{*}{$L_{47-69}$}   & LS    & 0.94 & 0.18 & 5.16 \\ \cline{2-5}
    & TLS     & 0.78 & 0.17 & 5.16 \\ \cline{2-5}
  & MTEE     & 0.64 & 0.16 & 3.37 \\ \cline{2-5}
 & EGLE & 0.15 & 0.14  & 2.04 \\ \hline
 \multirow{4}{*}{$L_{75-69}$}             & LS    & 0.74 & 0.08 & 7.25\\\cline{2-5}
       & TLS   & 0.74 & 0.08 & 6.93\\\cline{2-5}
 & MTEE     & 0.61 & 0.07 & 3.95 \\ \cline{2-5}
 & EGLE & 0.30 & 0.07  & 2.10 \\ \hline
\end{tabular}
\end{table}

\textcolor{black}{In the second set of simulations, the number of iterations required by MTEE and EGLE to reach the same level of tolerance ($\epsilon_1 = 0.0001$) was compared.
The initial guesses were kept the same. The results obtained are displayed in Table \ref{MTEE-time complexity comparison}. 
It is clear from the table that EGLE took fewer iterations
to attain the desired accuracy level. Furthermore, for the same number of samples used, EGLE took just under a second for each iteration (on a computer with 8 GB RAM and Intel $\mathrm{11^{th}}$ Gen Core i7 processor), while the MTEE method took approximately 30 seconds for each iteration.
The slowness of the MTEE method is due to the double summation involved in its implementation (see Appendix \ref{MTEE Overview}). 
Thus, it can be concluded from this empirical analysis that EGLE is computationally superior to the information theory-based MTEE method.
}
 
\begin{table}[H]
\color{black}
\centering
\caption{\textcolor{black}{Comparing number of iterations taken by MTEE and EGLE for four lines of the IEEE 118-bus system}}
\label{MTEE-time complexity comparison}
\begin{tabular}{|l|l|l|}
\hline
  & MTEE & EGLE \\ \hline
$L_{38-65}$ & 983 & 227            \\ \hline
$L_{8-9}$ & 1021 & 243         \\ \hline
$L_{47-69}$ & 1278 & 269               \\ \hline
$L_{75-69}$ & 1422 & 276            \\ \hline
\end{tabular}
\color{black}
\end{table}

\subsection{\textcolor{black}{Comparison of EGLE with constrained LS and TLS}}
\label{Comparison with advanced LS and TLS}
\textcolor{black}{In the previous sub-sections, the EGLE method was compared with the conventional (default) formulations of LS and TLS. However, it is possible to improve the accuracy of these classical methods by incorporating some physical properties of the TLPE problem into their default formulations. From Section \ref{TLPEAbstraction}, it can be realized that $Y_1 + Y_3 = 0$. This information can be incorporated into the LS and TLS formulations as an equality constraint. 
It was also described at the start of this section that the line parameters vary within a pre-specified range (namely, $\pm 30\%$). This information can be incorporated into the LS and TLS formulations as an inequality constraint. 
Based on the aforementioned information, a constrained LS and TLS formulation was created and its performance compared with EGLE for different lines of the IEEE 118-bus system.}

\textcolor{black}{During implementation, it was observed that the inequality constraint remained inactive for all the lines. 
Furthermore, the equality constraint primarily impacted the resistance estimates. This is because the susceptance parameter does not depend on $Y_1$ or $Y_3$ (see \eqref{eq4}). Similarly, $Y_4 \gg Y_1$ and $Y_4 \gg Y_3$, implying that $(2Y_4)^2 \gg (Y_1 - Y_3)^2$. This leads to the following approximation in \eqref{eq4}: $(Y_1 - Y_3)^2 + (2Y_4)^2 \approx (2Y_4)^2$. Consequently, the reactance parameter is also minimally impacted when the conventional formulation is replaced by the constrained formulation.
The $\mathrm{ARE}$ values of the improved resistance estimates obtained using the constrained LS and TLS methods are compared with the conventional LS, TLS, and the EGLE estimates in Table \ref{CLS-CTLS-comparison}. 
It was observed from the table that the resistance estimates of both LS and TLS were improved by adding the constraints, but they were still inferior to those obtained using EGLE.}

\begin{table}[H]
\caption{\color{black}{Comparison of resistance $\mathrm{ARE}$ (\%) of constrained LS and TLS with conventional LS, TLS, and EGLE}}
\centering
\label{CLS-CTLS-comparison}
\begin{tabular}{|l|c|c|c|c|c|}
\hline
\color{black}Line & \color{black}LS   & \color{black}TLS  & \color{black}Constrained LS  & \color{black}Constrained TLS & \color{black}EGLE \\ \hline
\color{black}$L_{38-65}$ & \color{black}0.77 & \color{black}0.69 & \color{black}0.38 & \color{black}0.32 & \color{black}0.08 \\ \hline
\color{black}$L_{8-9}$    & \color{black}0.49 & \color{black}0.45 & \color{black}0.25 & \color{black}0.20  & \color{black}0.04 \\ \hline
\color{black}$L_{47-69}$    & \color{black}0.94 & \color{black}0.78 & \color{black}0.49 & \color{black}0.41  & \color{black}0.08 \\ \hline
\color{black}$L_{75-69}$    & \color{black}0.74 & \color{black}0.74 & \color{black}0.35 & \color{black}0.35 & \color{black}0.24 \\ \hline
\end{tabular}
\end{table}


\section{Results: TLPE using Actual PMU Data}
\label{TLPE using real PMU data}
\textcolor{black}{
In this section, TLPE is carried out using raw PMU data obtained from a U.S. power utility in the Eastern interconnection.
This data was collected over a period of 3 consecutive years from about 400 PMUs that had been placed across the utility's service area.
To avoid ill-conditioning issues, the data was down sampled from 30 samples per second to 1 sample per minute.
The initial guesses for the line parameters were obtained from the PSS/E database shared by the utility, while a zero-mean Gaussian distribution was assumed as the initial guess for $c_e$ and $D_e$.
As the true line parameters are not known, $\mathrm{ARE}$ cannot be computed for comparison purposes for the analysis conducted in this section. 
Therefore, the ability of EGLE to track the line parameters is assessed by the consistency of the estimates across similar operating conditions, as described below.
}

\textcolor{black}{Two sets of PMU data 
corresponding to both the ends of a transmission line
were extracted from the massive database shared by the utility. The first set, $\mathrm{S1}$, comprised current and voltage phasor measurements from \textit{three} weekdays - Monday, Wednesday, and Friday - over a period of \textit{two} consecutive weeks. The second set, $\mathrm{S2}$, comprised current and voltage phasor measurements from \textit{two} weekdays - Tuesday and Thursday - over a period of \textit{three} consecutive weeks.
For both the sets, PMU data corresponding to two different time intervals were considered: the first time interval, denoted by $\mathrm{T1}$, was from 8 AM to 11 AM, while the second interval, denoted by $\mathrm{T2}$, was from 3 PM to 6 PM.
When the PMU data collected for the two sets and over the two time intervals were fed as inputs to EGLE, the results shown in Table \ref{OGnE-data-LP-consistency} were obtained.
}


\begin{table}[ht]
\color{black}
\centering
\caption{\textcolor{black}{Line parameter estimation using actual PMU data}}
\label{OGnE-data-LP-consistency}
\begin{tabular}{|l|c|c|c|c|c|c|}
\hline
    & \multicolumn{2}{c}{r (in p.u.)}        & \multicolumn{2}{|c|}{x (in p.u.)}                                     & \multicolumn{2}{c|}{b (in p.u.)}                               \\ \hline
   & T1      & T2      & T1                             & T2      & T1                            & T2        \\ \hline
S1 & 0.00396 & 0.00413 &  0.01950  & 0.01953 & 0.5088 & 0.5091    \\ \hline
S2 & 0.00392 & 0.00405 &  0.01947& 0.01951 & 0.5074 & 0.5096    \\ \hline
\end{tabular}
\color{black}
\end{table}

\textcolor{black}{From the table it is clear that for the same time intervals the parameter estimates are consistent across $\mathrm{S1}$ and $\mathrm{S2}$. This is expected because for similar ambient temperature and loading conditions, the line parameters are expected to be similar. 
When the estimates are compared across the two time intervals, slight variations are observed, particularly in the resistance estimates; this can be attributed to the difference in ambient temperatures for $\mathrm{T1}$ and $\mathrm{T2}$.
Thus, it can be concluded from this analysis that: (a) line parameters do change over time, and (b) EGLE is able to track the variations in the line parameters from actual PMU data.
}

\section{Discussion}
\label{Discussion}
\subsection{Generalization Capability of EGLE}
\label{Model selection for the proposed method}

\begin{figure}[ht]
            \centering
            \includegraphics[width=0.45\textwidth]{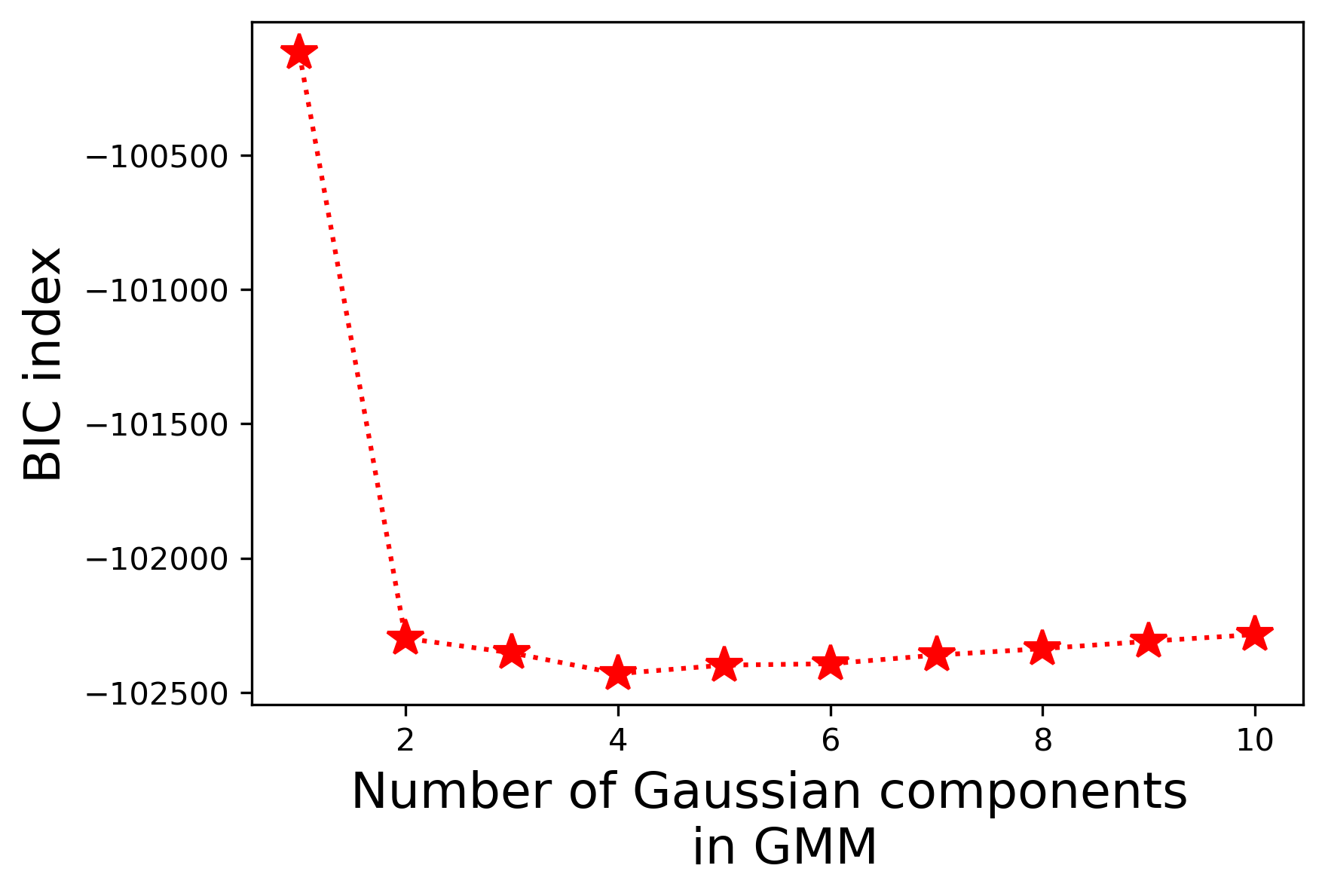}
            \caption{\textcolor{black}{Selecting optimal number of GMM components using BIC}}%
            \label{BIC_vs_nGC}
\end{figure}

\textcolor{black}{EGLE is expected to perform TLPE irrespective of the number of GMM components required to model the non-Gaussian noise in the PMU measurements.
To validate this expectation, a four-component GMM noise was added to the simulated voltage and current phasor measurements corresponding to $L_{38-65}$ of the IEEE 118-bus system.  
This noise model was characterized by a mean of ($\begin{bmatrix}
-0.002 &  0 & 0.005 & 0.008
\end{bmatrix}$), standard deviation of ($\begin{bmatrix}
  0.001 & 0.001  & 0.001  & 0.001
\end{bmatrix}$), and weight of ($ \begin{bmatrix}
 0.1 & 0.2 & 0.5 & 0.2
\end{bmatrix}$).
The noisy measurements were fed as inputs to EGLE, and the BIC values obtained for $m$ varying from $1$ to $10$ are shown in Fig. \ref{BIC_vs_nGC}. 
From the figure, it is clear that the optimal number of GMM components chosen by BIC is 4, which is same as the number of GMM components present in the noise.
The parameter estimates corresponding to $m=4$ are shown in Table \ref{sensitivity_toward_Number_of_Gaussian_components}. 
From the table it is evident that the line parameter estimates obtained using EGLE outperform those obtained using traditional estimation methods (LS and TLS).
Since the number of GMM components used to approximate the non-Gaussian noise was determined by EGLE and not provided as an \textit{a priori} knowledge, this analysis gives a good demonstration of its generalization capability.
}

\begin{table}[H]
\color{black}
\centering
\caption{\textcolor{black}{Parameter estimates for $L_{38-65}$ of IEEE 118-bus system for a $4$-component GMM noise in the measurements over 1,000 MC runs}}
\label{sensitivity_toward_Number_of_Gaussian_components}
\begin{tabular}{|l|l|l|l|}
\hline
 &
LS &
TLS&
EGLE \\ \hline
r-MARE(\%) & 1.42   &  1.38  & 0.3  \\
\hline
x-MARE(\%) & 0.36  & 0.36  & 0.25  \\
\hline
b-MARE(\%) & 2.31  & 2.26  & 1.10  \\
\hline
$\mathrm{MARE_{net}}$ (\%) & 2.74  & 2.67  & 1.17 \\
\hline
\end{tabular}
\color{black}
\end{table}

\subsection{Sensitivity Analysis}
\label{Results-Sensitivity analysis}

In this sub-section, we perform sensitivity analyses to further investigate the robustness 
of EGLE. 

\subsubsection{Impact of Measurement Noise Levels on Estimation}
To quantify the impact of measurement noise levels on TLPE, experiments were conducted on $L_{38-65}$, with GMM measurement noises ranging between $\pm 1 \%$, $\pm 2 \%$, $\pm 5 \%$,  and $\pm 10 \%$, respectively, in the PMU measurements.
This sensitivity study depicts a situation where there is increasing degradation in the instrumentation system of the PMUs providing the measurements.
Note that the GMM noise characteristics corresponding to $\pm 1 \%$ noise is similar to the GMM noise characteristics used in 
Section \ref{Results}.
For each successive noise level, the GMM noise characteristics was obtained by suitably scaling the mean and standard deviations of the noise.
For example, to obtain the GMM noise characteristics of $\pm 2 \%$ noise level, the GMM noise characteristics corresponding to $\pm 1 \%$ noise level was scaled by a factor of two. 

\begin{figure}[ht]
            \centering
            \includegraphics[width=0.40\textwidth]{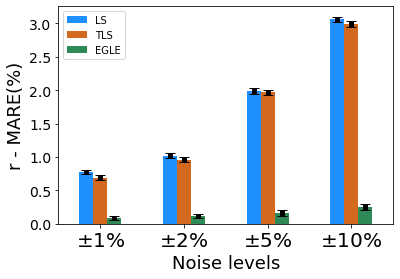}
            \caption{Impact of measurement noise levels on estimated resistance of $L_{38-65}$ of the IEEE 118-bus system}
            \label{TLPE_NoiselevelEffect_Resistance}
\end{figure}

\begin{figure}[ht]
            \centering
            \includegraphics[width=0.40\textwidth]{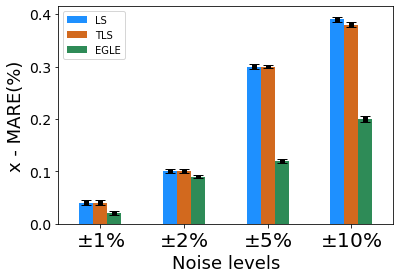}
            \caption{Impact of measurement noise levels on estimated reactance of $L_{38-65}$ of the IEEE 118-bus system}
            \label{TLPE_NoiselevelEffect_Reactance}
\end{figure}

\begin{figure}[ht]
            \centering
            \includegraphics[width=0.40\textwidth]{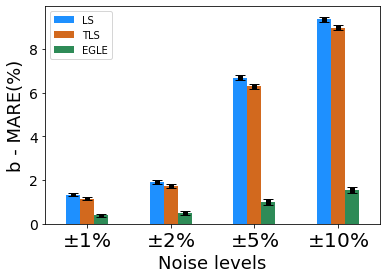}
            \caption{Impact of measurement noise levels on estimated susceptance of $L_{38-65}$ of the IEEE 118-bus system}
            \label{TLPE_NoiselevelEffect_Susceptance}
\end{figure}

\begin{figure}[ht]
            \centering
            \includegraphics[width=0.40\textwidth]{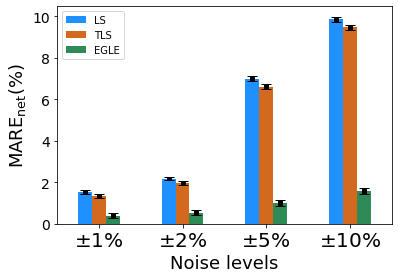}
            \caption{Impact of measurement noise levels on $\mathrm{MARE_{net}}$ of $L_{38-65}$ of the IEEE 118-bus system}
            \label{TLPE_NoiselevelEffect_NetARE}
\end{figure}

The $\mathrm{MARE}$ of the resistance, reactance, and shunt susceptance estimates obtained following 1,000 MC runs are displayed in \color{black}Fig. \ref{TLPE_NoiselevelEffect_Resistance}-\ref{TLPE_NoiselevelEffect_Susceptance} \color{black}. 
It can be noticed from the figures that the amount of noise present in the measurements has a considerable impact on the parameter estimates. 
Specifically, the susceptance parameter was observed to be highly prone to the inaccuracies present 
in the measurements. Conversely, the reactance has a relatively better tolerance towards noise content. 
For all four ranges of noise considered in this analysis, EGLE performed significantly better than LS and TLS for all three line parameters.
This observation is also reflected in the $\mathrm{MARE_{net}}$ shown in Fig. \ref{TLPE_NoiselevelEffect_NetARE}. 
The contrast in the performance was particularly glaring when the noise content was high ($\pm 10 \%$). 

\subsubsection{Impact of Initial Conditions on Estimation}
Being an iterative method, EGLE requires an initial guess of the parameters. 
The analysis conducted in \cite{kusic2004measurement, mansani2018estimation} reveals that the initial guess is expected to be within 30\% of the true value.
For studying the effect of initial condition on EGLE, a relative initialization (RI) index is defined as shown below,
\begin{equation}
    \mathrm{RI \: index} = \frac{|x^{(0)}-x^*|}{x^*}
\end{equation}
where, $x^{(0)}$ denotes the initial guess and $x^*$ denotes the true value of the parameter to be estimated.
The $\mathrm{MARE}$ observed over 1,000 MC runs for $L_{38-65}$ while starting with progressively worse initial guesses 
is shown in Table \ref{Effect_of_IG_PE}. It can be observed from the table that the estimates obtained using EGLE had similar accuracy for RI index less than $0.3$. 
The quality of the parameter estimates did deteriorate (and became worse than those obtained using LS and TLS) when the RI index increased over 0.4. This is not surprising considering the non-convex nature of the problem being solved in this paper. However, the chances of this happening in reality (i.e., initial guess being more than $\pm 40 \%$ away from the true value) is miniscule for the TLPE problem.  

\begin{table}[H]
\centering
\caption{Impact of \color{black}initial conditions \color{black}on estimating parameters of $L_{38-65}$ using EGLE {\label{Effect_of_IG_PE}}}
\begin{tabular}{|l|l|l|l|l|l|l|l|l|}
\cline{1-6}
   &\multicolumn{3}{c|}{EGLE}& \multirow{2}{*}{LS}   & \multirow{2}{*}{TLS}   \\ \cline{1-4}
RI index     &$[0,0.1]$ & $[0.1,0.2]$    & $[0.2, 0.3]$   &   &   \\ \hline
r-MARE(\%) & 0.07  & 0.07 & 0.08  & 0.77 & 0.66 \\ \hline
x-MARE(\%) & 0.03  & 0.03 & 0.03  & 0.04 & 0.04 \\ \hline
b-MARE(\%) & 0.37  & 0.38 & 0.38  & 1.33 & 1.14 \\ \hline
\end{tabular}
\end{table}

\subsection{\textcolor{black}{Non-degeneracy of EGLE in Presence of Gaussian Noise and Measurement Bias}}

\begin{figure}[ht]
        \centering
            \begin{subfigure}[b]{0.4\textwidth}
            \centering 
            \includegraphics[width=\textwidth]{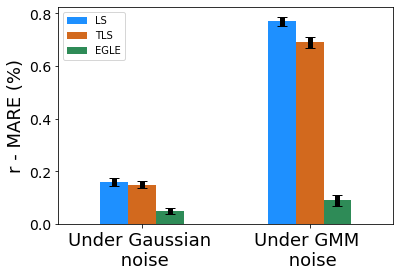}
            \vspace{-7mm}
            \caption[]%
            {{Resistance estimates
            }}    
              \vspace{1mm}
        \end{subfigure}
        \hfill
        \begin{subfigure}[b]{0.4\textwidth}
            \centering
            \includegraphics[width=\textwidth]{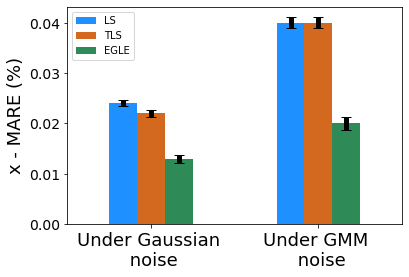}
             \vspace{-7mm}
            \caption[Real(Vdiff)]%
            {{Reactance estimates
            }}    
            \vspace{1mm}
        \end{subfigure}
         \hfill
        \begin{subfigure}[b]{0.4\textwidth}  
            \centering 
            \includegraphics[width=\textwidth]{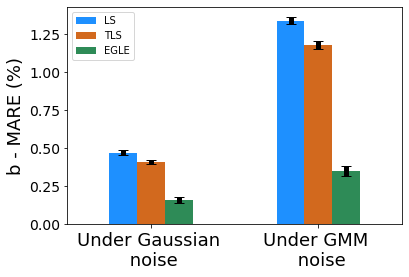}
             \vspace{-7mm}
            \caption[]%
            {{Susceptance estimates
            }}    
            \vspace{1mm}
        \end{subfigure}
        \caption{\textcolor{black}{Performance comparison in presence of Gaussian and non-Gaussian measurement noise for $L_{38-65}$ of the IEEE 118-bus system over 1,000 MC runs}}
        \label{GaussianVsNonGaussian}
\end{figure}

\textcolor{black}{In this sub-section, we study the case where the PMU measurements have Gaussian noise.
As Gaussian is a special case for a GMM (namely, that the mixtures are identical), we compare performance of EGLE with that of LS and TLS.
A two component GMM having the characteristics defined in Section \ref{Results-GMM Noise in Dependent variables} is used to model the non-Gaussian noise in the voltage and current phasor measurements.
For ensuring that the Gaussian noise has a similar same range/spread (as the non-Gaussian noise), its mean was kept at $0.0035$ and its standard deviation was kept at $0.0027$.
Hence, this simulation also investigates the case where a \textit{bias} is present in the PMU measurements \cite{ahmad2019statistical, salls2021statistical}. 
The analysis was performed for $L_{38-65}$ of the IEEE 118-bus system.
The comparison of the $\mathrm{MARE}$ for the line parameter estimates is shown in Fig.  \ref{GaussianVsNonGaussian}.}

\textcolor{black}{It can be observed from the figure that the $\mathrm{MARE}$ for the traditional approaches (LS and TLS) increases considerably when the measurement noise model changes from Gaussian to non-Gaussian (compare the heights of the blue and orange bars of the resistance, reactance, and susceptance estimates). However, the increase is minor for EGLE (the height of the green bars only change slightly when the noise model is changed). 
The slight improvement in the performance of EGLE over LS and TLS in the presence of Gaussian noise (compare heights of blue and orange bars with the green bars for the Gaussian noise case), is because of the former's ability to better handle bias in the measurements.
To summarize, this analysis shows that for operating conditions in which the measurement noise is non-Gaussian, EGLE significantly outperforms the traditional methods. Even when the measurement noise is Gaussian but has a bias (which often occurs in PMU measurements \cite{ahmad2019statistical, salls2021statistical}), EGLE is still able to give better estimates.}

\subsection{Computational Complexity Analysis}
\label{Computational Complexity Analysis}
\textcolor{black}{The computational complexity of EGLE and the techniques used for comparison are described in this sub-section in terms of the big O notation (denoted by $\mathcal{O}$).
The computational complexity of the four techniques that were used in the simulations are as follows: (a) For a linear regression problem with $N$ number of samples,
the LS method has a computational complexity of $\mathcal{O}(N)$; note that as the number of parameters is constant for the TLPE problem, it was not included in the computational complexity calculation. (b) If the TLS method is implemented using the truncated SVD approach, then it also has a computational complexity of $\mathcal{O}(N)$ \cite{diao2019total}.
(c) The denoising-followed-by-(conventional)-estimation approach also has a computational complexity of $\mathcal{O}(N)$. (d) If $\beta$ denotes the maximum number of iterations, then the computational complexity of the MTEE approach is $\mathcal{O}(\beta N^2)$  \cite{shen2015minimum}. (e) The computational complexity of EGLE after incorporating the BIC-based model selection is $\mathcal{O}(\beta N)$. 
Thus, it can be inferred from this comparison that the computational complexity of EGLE is greater than the LS, TLS, and denoising-based methods but much less than the MTEE method.
}

\section{Conclusion}
A novel method (termed EGLE) for jointly estimating accurate transmission line parameters and noise parameters when the PMU measurements have non-Gaussian measurement noises has been developed and presented in this paper. The effectiveness of EGLE for TLPE was compared with that of the LS, TLS, \textcolor{black}{denoising-based, and MTEE} methods. The results show that EGLE significantly outperform the traditional methods 
when the measurement noise is highly non-Gaussian. It was also shown that 
if the measurement noise is Gaussian (a special case of GMM), EGLE
continues to give accurate estimates.
Thus, EGLE is a more general method for parameter estimation that can be used for any type of measurement noise. 
The ability to estimate the non-Gaussian noise characteristics (by expressing them as a GMM) is an added advantage of the proposed methodology.

Accurate knowledge of the transmission line parameters is crucial for improved power system monitoring, control, and protection applications. 
As the proliferation of PMUs increases in the power system (from high voltage transmission to sub-transmission and even distribution), the use of the proposed methodology will ensure that monitoring, control and protection applications at any level of the power system is not negatively impacted by inaccurate line parameter information.

\appendices
\section{\textcolor{black}{Nature and Source of Noise in {PMU} Measurements}}
\label{AppendixA}
\textcolor{black}{
Most papers published in the literature on PMUs have implicitly assumed that the noise in the synchrophasor measurement system has a Gaussian distribution. It is only recently that extensive statistical testing conducted on data obtained from PMUs placed in the field has proven otherwise. One of the first studies was conducted by Wang et al. in \cite{wang2017assessing}. They used nine sets of redundant PMU measurements from 18 buses of the Western Electricity Coordinating Council system. The conclusion of their statistical analysis was that noise in the PMU measurements \textit{did not} follow a Gaussian distribution. An independent study was conducted by Ahmad et al. \cite{ahmad2019statistical} using field PMU data from the Texas Independent Synchrophasor network and the Indian synchrophasor network. The conclusion of their statistical analysis was that for a given window, a GMM is appropriate for modeling the noise in synchrophasor measurements. These two studies along with \cite{salls2021statistical, saadedeen2021gps} have identified the following to be the source of noise in PMU measurements: different system operating conditions, aging process of instrument transformers, incorrect time synchronization, errors introduced by the phasor estimation algorithm, varying communication channel noises, and/or cyber-attacks such as eavesdropping, global positioning system (GPS) spoofing and data tampering.}

\section{\textcolor{black}{Approximating an arbitrary distribution by a Gaussian Mixture Model (GMM)}}
\label{AppendixB}
\renewcommand{\theequation}{\thesection.\arabic{equation}}
\setcounter{equation}{0}
\textcolor{black}{GMMs are a powerful way of representing any non-Gaussian density with sufficient accuracy.
This can be mathematically shown using the properties of a delta function \cite{plataniotis2017gaussian}. 
A family of functions, $\delta_{\lambda}$, on the interval ($-\infty, \infty$) which are integrable over every interval are called a delta family of positive types if
\begin{itemize}
    \item $\int_{-a}^a \delta_{\lambda}(x) dx \xrightarrow[]{} \lambda, \text{ as } \lambda \xrightarrow[]{} \lambda_0,$ for some $a$.
    \item For every constant $\gamma > 0$, $\delta_{\lambda}$ tends to zero uniformly for $\gamma \leq |x| \leq \infty$ as $\lambda \xrightarrow[]{} \lambda_0$.
    \item $\delta_{\lambda}(x) \geq 0 $ for all $x$ and $\lambda$
\end{itemize}
}

\textcolor{black}{Additionally, note that when the variance tends to $0$, the Gaussian density tends to the delta function.
Now, let us look at approximating an arbitrary function $p$ using the delta family. Consider the sequence $p_{\lambda}(x)$, which is formed by the convolution of $\delta_{\lambda}$ and $p$, given by
\begin{equation}
    p_{\lambda}(x) = \int_{-\infty}^{\infty} \delta_{\lambda}(x-u) p(u) du. 
\end{equation}
It can be observed that $p_{\lambda}(x)$ converges to $p(x)$ on every interior sub-interval of $(-\infty, \infty)$.
Since the Gaussian density can be used as a delta family of positive type, the approximation $p_{\lambda}$ can be written as:
\begin{equation}
    p_{\lambda}(x) = \int_{-\infty}^{\infty} \mathcal{N}_{\lambda}(x-u) p(u) du. 
\end{equation}
This forms the basis for the Gaussian sum approximation. $p_{\lambda}(x)$ can be approximated on any finite interval by a Riemann sum, since the term $\delta_{\lambda}(x-u) p(u)$ is integrable on ($-\infty, \infty$) and is at least piece-wise continuous. If a bounded interval $(a, b)$ is considered, this function is given by:
\begin{equation}
    p_{\lambda, n}(x) = \frac{1}{k} \sum_{i=1}^n \mathcal{N}_{\lambda}(x-x_i) [\xi_i - \xi_{i-1}] 
\end{equation}
where the interval $(a,b)$ is divided into $n$ sub-intervals by selecting points such that:
\begin{equation}
    a=\xi_0 < \xi_1 < \xi_2 < \dots < \xi_n = b.
\end{equation}
Using the mean value theorem, in each sub-interval, a point $x_i$ can be chosen such that: 
\begin{equation}
   p(x_i)[\xi_i - \xi_{i-1}] = \int_{-\xi_i - 1}^{\xi_i} p(x) dx
\end{equation}
Thus, an approximation of $p_{\lambda}$ over some bounded interval $(a,b)$ can be written as:
\begin{equation}
    p_{\lambda, n}(x) =  \sum_{i=1}^n \omega_i \mathcal{N}_{\sigma_i}(x-x_i) 
\end{equation}
where $\sum_{i=1}^n \omega_i = 1$ and $\omega \geq 0 \: \forall \: i$. \\
Under this framework, an unknown $d$-dimensional distribution can be expressed as a linear combination of Gaussian terms. The form of the approximation becomes:
\begin{equation}
    p(x) =  \sum_{i=1}^{m} \omega_i \mathcal{N}(x; \mu_i, \Sigma_i)
\end{equation}
where, $m$ denotes the number of Gaussian components required to approximate the non-Gaussian distribution in the form of a GMM, $\omega_i$ is the weight of the $i^{th}$ Gaussian component, $\mathcal{N}(x; \mu_i, \Sigma_i)$ denotes the $i^{th}$ Gaussian component given by $\mathcal{N}(x; \mu_i, \Sigma_i) = \frac{1}{(2 \pi)^{0.5} |\Sigma_i|^{0.5}} exp(-0.5(x-\mu_i)^T \Sigma_i^{-1}(x-\mu_i))$, and $\mu_i$ and $\Sigma_i$ denotes the mean and covariance matrix of the $i^{th}$ Gaussian components.\\}

\section{\textcolor{black}{Overview of Minimum Total Error Entropy (MTEE) Method}}
\label{MTEE Overview}
\setcounter{equation}{0}
\textcolor{black}{An alternate way to estimate parameters for a static linear regression problem in which both the dependent and the independent variables have non-Gaussian noise is by minimizing the total error entropy. This was done in \cite{shen2015minimum}, and the resulting technique was referred to as the minimum total error entropy (MTEE) method. The total error was defined as
\begin{equation}
    \begin{aligned}
        e^{tot} = \frac{c - D x}{\sqrt{x^T x + 1}}
    \end{aligned}
\end{equation}
The MTEE method minimized the quadratic Renyi’s entropy of $e^{tot}$. 
Using the Parzen window method, an expression for the quadratic Renyi's entropy was obtained as follows
\begin{equation}
    \begin{aligned}
        \hat{H}_2(e^{tot}) = -\mathrm{log}\left(\frac{1}{N^2} \sum_{i=1}^N \sum_{j=1}^N G_{\sigma\sqrt{2}} \left(e_j^{tot} - e_i^{tot}\right)\right)
    \end{aligned}
    \label{MTEE_minimization}
\end{equation}
where $N$ is the length of the Parzen window, and $G_\sigma (.)$ is a Gaussian kernel with kernel size $\sigma$. In \cite{shen2015minimum}, the minimization of $\hat{H}_2(e^{tot})$ was performed iteratively using the steepest descent method.
Although the MTEE method is able to estimate parameters for EIV problems in which the noises in the dependent and the independent variables are non-Gaussian, it takes a long time to converge because of the double summation over $N$ present in \eqref{MTEE_minimization}.
}

\bibliographystyle{IEEEtran}

\bibliography{References/TLPE_PS_Ref_2,References/TLS_Adj_Ref,References/Non-Gaussian_measurement_noise_References, References/TLPE_recent_references}




%








\end{document}